# Efficient Large-Scale Graph Processing on Hybrid CPU and GPU Systems

ABDULLAH GHARAIBEH, The University of British Columbia
TAHSIN REZA, The University of British Columbia
ELIZEU SANTOS-NETO, The University of British Columbia
LAURO BELTRÃO COSTA, The University of British Columbia
SCOTT SALLINEN, The University of British Columbia
MATEI RIPEANU, The University of British Columbia

The increasing scale and wealth of inter-connected data, such as those accrued by social network applications, demand the design of new techniques and platforms to efficiently derive actionable knowledge from large-scale graphs. Large real-world graphs, however, are famously difficult to process efficiently. Not only do they have a large memory footprint, but most graph algorithms also entail memory access patterns with poor locality, data-dependent parallelism, and a low compute-to-memory access ratio. To complicate matters further, many real-world graphs have a highly heterogeneous node degree distribution (i.e., they are scale-free), hence partitioning these graphs for parallel processing and simultaneously achieving access locality and load-balancing is difficult if not impossible.

This work starts from the hypothesis that hybrid platforms (e.g., GPU-accelerated systems) have both the potential to cope with the heterogeneous structure of scale-free graphs and to offer a cost-effective platform for high-performance graph processing. This work assesses this hypothesis and presents an extensive exploration of the opportunity to harness hybrid systems to process large-scale scale-free graphs efficiently. In particular, *(i)* we present a performance model that estimates the achievable performance on hybrid platforms; *(ii)* informed by the performance model, we design and develop TOTEM – a processing engine that provides a convenient environment to implement graph algorithms on hybrid platforms; *(iii)* we show that further significant performance gains can be extracted using partitioning strategies that aim to produce partitions that each matches the strengths of the processing element it is allocated to, and finally, *(iv)* we demonstrate the performance advantages of the hybrid system through a comprehensive evaluation that uses real and synthetic scale-free workloads (as large as 16 billion edges), multiple graph algorithms that stress the system in various ways, and a variety of hardware configurations.

Categories and Subject Descriptors: **C.1.3 [Processor Architectures]**: Other Architecture Styles – *Heterogeneous (hybrid) systems*. **G.2.2 [Discrete Mathematics]**: Graph Theory – *Graph Algorithms*

## 1. INTRODUCTION

Graphs are the core data structure for problems that span a wide set of domains, from mining social networks, to genomics, to business and information analytics. In these domains, key to our ability to transform raw data into insights and actionable knowledge is the capability to process large graphs efficiently and at a reasonable cost.

A major challenge when processing large graphs is their memory footprint: efficient graph processing requires the whole graph to be present in memory, and large real graphs can occupy gigabytes to terabytes of space. For example, a snapshot of the current Twitter follower network has over 500 million vertices and 100 billion edges, and requires at least 0.5TB of memory. As a result, the most commonly adopted solution to cost-efficiently process large-scale graphs is to partition them and use shared-nothing cluster systems [Malewicz et al. 2010; Gonzalez et al. 2012].

We observe, however, that today more efficient solutions are affordable: it is feasible to assemble single-node[1] platforms that aggregate 100s of GB to TBs of RAM and massive computing power [Gupta et al. 2013; Rowstron et al. 2012; Shun and Blelloch 2013] all from commodity components and for a relatively low budget. Compared to clusters, single-node platforms are easier to program, and promise better performance and energy efficiency for a large class of real-world graph problems. In fact, such single-node graph processing platforms are currently being used in production: for example, Twitter's 'Who To Follow' (WTF) service, which uses the follower network to recommend connections to users, is deployed on a single node [Gupta et al. 2013].

Despite these recent advances, single-node platforms still face a number of performance challenges. First, graph algorithms have low compute-to-memory access ratio, which exposes fetching/updating the state of vertices (or edges) as the major overhead. Second, graph processing exhibits irregular and data-dependent memory access patterns, which lead to poor memory locality and reduce the effectiveness of caches and pre-fetching mechanisms. Finally, many real-world graphs have a highly heterogeneous node degree distribution (i.e., they have power-law degree distribution and are commonly named "scale-free") [Barabási 2003; Barabási et al. 2000; Jeong et al. 2001; Iori et al. 2008], which makes dividing the work among threads for access locality and load-balancing difficult.

In this context, two reasons (detailed in §2) support the intuition that commodity single-node hybrid systems (e.g., GPU-accelerated nodes) may be an appealing platform for high-performance, low-cost graph processing: First, Graphical Processing Units (GPUs) bring massive hardware multithreading able to mask memory access latency – the major barrier to performance for this class of problems. Second, a hybrid system that hosts processing units optimized for fast sequential processing and units optimized for bulk processing matches well the heterogeneous structure of the many graphs that need to be processed in practice.

This paper investigates these premises. More precisely, *it investigates the feasibility and the comparative advantages of supporting graph processing of scale-free graphs on hybrid, GPU-accelerated nodes*.

***The following high-level questions guide our investigation:***

*Q1. Is it feasible to efficiently combine traditional CPU cores and massively parallel processors (e.g., GPUs) for graph processing?* In particular, what are the general challenges to support graph processing on a single-node GPU-accelerated system?

*Q2. How should the graph be partitioned to efficiently use both traditional CPU cores and GPU(s)?* More specifically, are there low-complexity partitioning algorithms that generate a workload allocation that matches well the individual strengths of CPUs and GPUs?

Making progress on answering these questions is important in the context of current hardware trends: as the relative cost of energy continues to increase relative to the cost of silicon, future systems will host a wealth of different processing units. In this context, partitioning the workload and assigning the partitions to the processing element where they can be executed most efficiently in terms of power or time becomes a key issue.

***Contributions.*** This work demonstrates that partitioning large scale-free graphs to be processed concurrently on hybrid CPU and GPU platforms offers significant performance gains (we have demonstrated [Gharaibeh et al. 2013b] that these gains hold for energy as well). Moreover, this work defines the class of partitioning algorithms that will enable best performance on hybrid platforms: these algorithms should focus on shaping the workload to best match the bottleneck processing engine, rather than on minimizing communication overheads. Finally, we experiment with a few partitioning solutions from this class, analyze the observed performance, and propose guidelines for when they should be used.

---

[1] We use *node* to refer to processing elements (i.e., machines), and *vertex* to refer to the graph element.

In more detail, the contributions are:
- *A performance model* (§3) to assess the feasibility of accelerating large-scale graph processing by offloading a graph partition to the GPU. The model is agnostic to the exact graph processing algorithm, and it takes into account only a small number of key aspects such as the parallel processing model, the characteristics of the processing elements, and the properties of the communication channel among these elements. The model supports the intuition that keeping the communication overhead low is crucial for efficient graph processing on hybrid systems and it prompts us to explore the benefits of message reduction to minimize these overheads.
- *TOTEM[2]: an open-source graph processing engine for GPU-accelerated platforms* (§4). TOTEM enables efficiently using all CPU and GPU cores on a given node all while limiting the development complexity. TOTEM offers a number of important functionalities such as: supporting a BSP parallel programming model, ghost nodes to handle boundary edges, graph representation for both the host and accelerator, and graph partitioning techniques to mention a few. Additionally, guided by the performance model, TOTEM embeds a set of optimizations key to achieve the desired performance levels: message aggregation and reduction, transparent use of mapped memory, and overlapping communication with computation.
- *Insights into key performance overheads* (§5). Using TOTEM's abstractions, we implement four graph processing algorithms that stress the hybrid system in different ways. We demonstrate that the gains predicted by the model are achievable in practice when offloading a random partition to the GPUs. Moreover, we show that the optimizations applied by TOTEM significantly reduce the overheads to communicate among the processing elements, and that the computation phase becomes the dominating overhead.
- *Low-cost partitioning strategies tailored for processing scale-free graphs on hybrid systems* (§6). Since the optimizations we apply eliminate communication as a major bottleneck, we focus on partitioning strategies that aim to reduce the computation bottleneck. These strategies aim utilize the heterogeneity in scale-free graphs to produce partitions such that the workload assigned to the bottleneck processing element exploits well the element's strengths. Our partitioning strategies are informed by vertex-connectivity, and lead to super-linear performance gains with respect to the share of the workload offloaded to the GPUs.
- *Application evaluation* (§7, §8, and §9.4). We demonstrate that the gains offered by the hybrid system hold for key applications: ranking web pages using PageRank, finding the main actors in a social network using Betweenness Centrality algorithm, and computing point-to-point shortest paths in a network using Single Source Shortest Path algorithm using large real-world graphs with over 3 billion edges (§7). We evaluate scalability on synthetic graphs with up to 16 billion edges in §8. Using five graph algorithms (breadth-first search, connected components in addition to the three algorithms mentioned above), we favorably compare with other platforms including Galois, Ligra and PowerGraph (§9.4): TOTEM, deployed on a modest one CPU socket and one GPU hybrid system, is more than 2x higher compared to the best performance achieved by state-of-the-art frameworks on a shared-memory machine with four high-end CPU sockets.

---

[2] The code can be found at: http://netsyslab.ece.ubc.ca

## 2. GRAPH PROCESSING ON HYBRID PLATFORMS: OPPORTUNITIES AND CHALLENGES

The previous section discussed the general challenges of single-node graph processing. This section details the opportunities and challenges brought by GPU acceleration in this context.

*The opportunities:* GPU-acceleration has the potential to offer the key advantage of massive, hardware-supported multithreading. In fact, current GPUs not only have much higher memory bandwidth than traditional CPU processors, but can mask memory access latency as they support orders of magnitude more in-flight memory requests through hardware multithreading.

Additionally, properly mapping the graph-layout and the algorithmic tasks between the CPU(s) and the GPU(s) holds the promise to exercise each of these computing units where they perform best: CPUs for fast sequential processing and GPUs for the bulk parallel processing.

In particular, this work focuses on harnessing the heterogeneity of vertex degree distribution in scale-free graphs. For example, the few high-degree vertices can be processed by the CPU, while the many low-degree ones can be processed on the GPU. While this limits the scope of this work, it still benefits various high-impact applications as many real-world graphs are scale-free. Examples of such graphs include social networks [Kwak et al. 2010], the Internet [Faloutsos et al. 1999], the World Wide Web [Barabási et al. 2000], financial networks [Iori et al. 2008], protein-protein interaction networks [Jeong et al. 2001], and airline networks [Wang and Chen 2003] to mention few.

*The challenges:* Large-scale graph processing poses two major challenges to hybrid systems. First, the large amount of data to be processed and the need to communicate between processors put pressure on two scarce resources: the GPUs' on-board memory and the host-to-GPU transfer bandwidth. Intelligent graph representation, partitioning and allocation to compute elements are key to reduce memory pressure, limit the generated PCI bus transfer traffic, and efficiently harness each processing element in an asymmetrical platform.

Second, to achieve good performance on GPUs, the application must, as much as possible, match the SIMD computing model. As graph problems exhibit data-dependent parallelism, traditional implementations of graph algorithms lead to low memory access locality. Nevertheless, GPUs are able to hide memory access latency via massive hardware multithreading that, with careful design of the graph data structure and thread assignment, can reduce the impact of these factors.

Finally, there is the additional challenge of mapping high-level abstractions (e.g., vertex-centric processing) and APIs to facilitate application development to the low-level infrastructure while limiting the efficiency loss.

## 3. MODELING HYBRID SYSTEMS' PERFORMANCE

Our model aims to provide insights to answer the following question: *Is it beneficial to partition the graph and process it in parallel on both the host and the GPU instead of processing it on the host only?*

It is worth stressing that our goal is a simple model that captures the key characteristics of a GPU-accelerated platform, highlights its bottlenecks, and helps reason about the feasibility of offloading. We deliberately steer away from a complex (though potentially more accurate) models. Our evaluation validates that this choice provides accuracy high-enough to lead to useful conclusions.

## 3.1 Notations and Assumptions

Let $G = (V, E)$ be a directed graph, where $V$ is the set of vertices and $E$ is the set of directed edges; $|V|$ and $|E|$ represent their respective cardinality. Also, let $P = \{p_{cpu}, p_{gpu}\}$ be the set of processing elements of a hybrid node (Figure 1). While the model can be easily generalized to a mix of multiple CPUs and GPUs; for the sake of simplicity, here we use a setup with only two processing units.

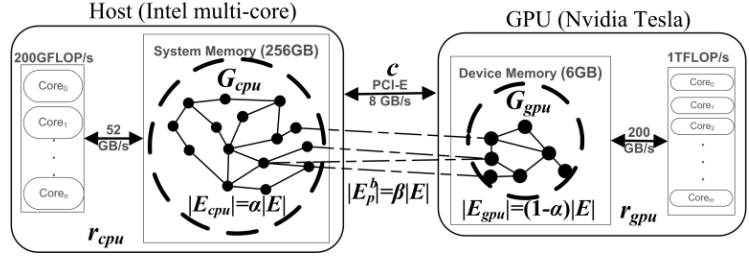

**Figure 1: An illustration of the model, its parameters, and their values for today's state-of-the-art commodity components.**

| | |
|---|---|
| $r_{cpu}$ $r_{gpu}$ | Processing rates on the CPU and GPU |
| $c$ | Communication rate between the host and GPU |
| $\alpha$ | Ratio of the graph edges that remain on the host |
| $\beta$ | Ratio of edges that cross the partition |

The model makes the following assumptions:

*(i)* Each processing element has its own local memory. The processing elements are connected by a bidirectional interconnect with communication rate $c$ measured in edges per second (E/s) – this is a reasonable unit as the time complexity of a large number of graph algorithms depends on the number of edges in the graph. The same model, however, can be recast in terms of vertex-centric algorithms by normalizing by the number of vertices instead of edges.

*(ii)* Once the graph is partitioned, the GPU processes its partition faster. This is because: first, GPUs have a higher graph processing rate than CPUs (based on published results [Hong et al. 2011a; Hong et al. 2011b], which we validated independently); second, GPUs have significantly less memory than the host, which limits the size of the offloaded partition.

*(iii)* The model assumes the overheads of scheduling the workload (e.g., partitioning the graph) and gathering the results produced by each processor are negligible compared to the algorithm's processing time.

## 3.2 The Model

Under the assumptions stated in the previous section, the time to process a partition of $G$, $G_p = (V_p, E_p) \subseteq G$ on a processing element $p$ is given by:

$$t(G_p) = \frac{|E_p^b|}{c} + \frac{|E_p|}{r_p} \quad (1)$$

where $r_p$ is the processing rate of processor $p$ (in edges/s), and $E_p^b \subseteq E_p$ represents the subset of *boundary edges* – edges where either the source or the destination vertex is not located in $p$'s local memory.

Equation 1 estimates the time required to process a partition as a combination of the time it takes to communicate possible updates through boundary edges (communication phase) plus the time it takes to process the edges in that given partition on processor $p$ (computation phase). Intuitively, the higher the processing rate of a processing element, the lower is the processing time. Similarly, the less communication a processing element needs to access the edges in its partition, the lower is the processing time.

Now, we build on Equation 1 and define the makespan[3] of a graph workload $G$ on a given platform $P$ as follows:

$$m_P(G) = \max_{p \in P} \{t(G_p)\} \quad (2)$$

The intuition behind Equation 2 is that the performance of a parallel system is limited by its slowest component. Since, as discussed before, the model assumes that the host processes its partition slower than the GPU (assumption *ii*), the time spent on processing the CPU partition is always higher than that of the GPU partition (i.e., $t(G_{cpu}) > t(G_{gpu})$).

Hence, the speedup of processing a graph on a hybrid platform (compared to processing it on the host only) can be calculated by Equation 3, as follows:

$$s_P(G) = \frac{t_{\{cpu\}}(G)}{m_P(G)} = \frac{t_{\{cpu\}}(G)}{t_{\{cpu\}}(G_{cpu})} = \frac{|E|/r_{cpu}}{|E^b_{cpu}|/c + |E_{cpu}|/r_{cpu}} \quad (3)$$

To understand the gains resulted from moving a portion of the graph to the GPU, we rewrite Equation 3 by introducing two parameters that characterize the '*quality*' of the graph partition. Let $\alpha$ be the share of edges (out of the total number of graph edges $|E|$) that are assigned to remain on the host, similarly let $\beta$ be the percentage of *boundary* edges (i.e., the edges that cross the partition). Introducing these parameters, we have:

$$s_P(G) = \frac{|E|/r_{cpu}}{\beta|E|/c + \alpha|E|/r_{cpu}} = \frac{c}{\beta r_{cpu} + \alpha c} = \frac{1}{\frac{\beta \cdot r_{cpu}}{c} + \alpha} \quad (4)$$

As expected, Equation 4 predicts that a high host-accelerator interconnect communication rate ($c$) improves the speedup. In fact, if $c$ is set to infinity, the speedup can be approximated as $1/\alpha$. This is intuitive, as in this case the communication overhead becomes negligible compared to the time spent on processing the CPU's share of edges, and the speedup becomes proportional with the offloaded portion of the graph.

### 3.3 Setting the Model's Parameters

Figure 1 presents an illustration of the model with reasonable values for its parameters for a state-of-the-art commodity hybrid platform. We discuss them in turn:

- *Communication rate* (*c*) is directly proportional to the interconnect bandwidth and inversely proportional to the amount of data transferred per edge. The GPU is typically connected to the host via a PCI-E bus. Latest GPU models support PCI-E gen3.0, which has a measured transfer bandwidth of 12GB/sec. If we assume the data transferred per edge is a 4-byte value (e.g., the "*distance*" in Breadth-first Search or the "*rank*" in PageRank), the transfer rate *c* becomes 3 Billion E/s – or BE/s.

---

[3] Makespan: the time difference between the start and finish of a sequence of graph processing tasks [Pinedo 2012].

- *CPU's processing rate ($r_{cpu}$) depends on the CPU's characteristics, the graph algorithm and implementation, and the graph topology. We assume that a CPU-only implementation is available and can be run on the machine to obtain $r_{cpu}$. This is a reasonable assumption as one typically starts off by implementing a CPU version of the algorithm.

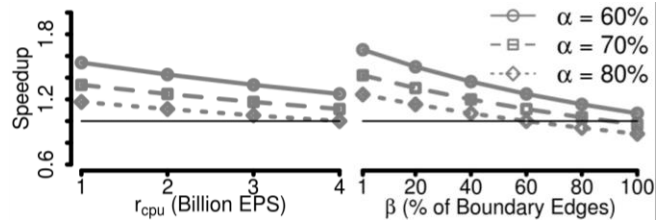

**Figure 2: Predicted speedup (values below one indicate slowdown). Left: while varying the CPU's processing rate ($\beta$ is set to 5%). Right: while varying the percentage of boundary edges ($r_{cpu}$ is set to 1 BE/s). The communication rate is 3 BE/s.**

- *Percentage of boundary edges ($\beta$)* depends on the graph partitioning between the processing elements. In the worst case, all edges cross the partition. Random partitioning leads to an average $\beta$=50%.
- *The share of the graph that stays on the CPU ($\alpha$)* is configurable, but is constrained by the memory space available on the processing elements. For example, larger memory on the GPU allows for offloading a larger partition, hence a smaller $\alpha$.

Figure 2 shows the speedup predicted by the model (Equation 4) for different values of $\alpha$, while varying the CPU processing rate (left plot) and the percentage of boundary edges (right plot). The values used for the CPU processing rate are informed by the best reported graph processing rates in the literature [Nguyen et al. 2013] for state-of-the-art commodity single-node machines.

The figure indicates that as the CPU processing rate increases (higher $r_{cpu}$, left plot) or for a graph partition that leads to larger percentage of boundary edges (higher $\beta$, right plot), the speedup decreases. This is because the communication overhead becomes more significant.

Nonetheless, the figure indicates that offloading part of the graph to be processed in parallel on the GPU can be beneficial. In particular, if $\beta$ is kept low (below 40% in Figure 2 (right)), the model predicts speedups. The figure also presents a hypothetical worst case where all of the edges are boundary edges (e.g., a bipartite graph where the partition cuts each edge). Even in this case, and due to the high communication rate $c$, a slowdown is predicted only for $\alpha > 70$%.

Finally, Figure 3 demonstrates the effect of the amount of transferred data per edge on the predicted speedup. As expected, the speedup drops as we double the amount of transferred data. However, if $\beta$ is kept low, the model predicts tangible speedups even when tripling the size of data transferred per boundary edge. To this end, the next section discusses how to keep $\beta$ low for scale-free graphs, the focus of this work.

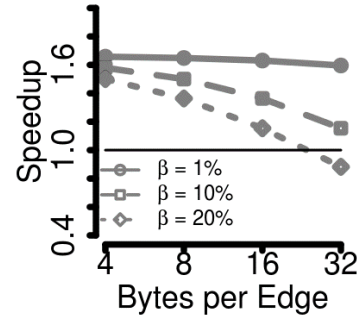

**Figure 3: Predicted speedup while varying the volume of transferred data per edge ($\alpha$ is set to 60% and $r_{cpu}$ to 1 BE/s).**

### 3.4 Reducing the Impact of Boundary Edges

This section presents an efficient technique that minimizes $\beta$, i.e., the percentage of boundary edges for scale-free graphs and a range of graph algorithms.

In particular, we explore the opportunity to reduce messages sent from multiple vertices residing in one processing element to a single vertex residing on the other. The intuition behind this optimization is that the power-law nature of scale-free graphs leads to a topology where multiple edges from the same partition point to the high-degree vertices on the other partition and thus enable message reduction.

Note that reduction is employed in other cluster-based graph processing frameworks [Malewicz et al. 2010] as well to reduce the communication overhead between partitions residing in different nodes. However, this technique is even more effective in the single hybrid node platform we target because the number of partitions we expect to have (e.g., two for a system with one GPU) is significantly lower than in the case of a distributed system with hundreds of compute nodes (hundreds of partitions).

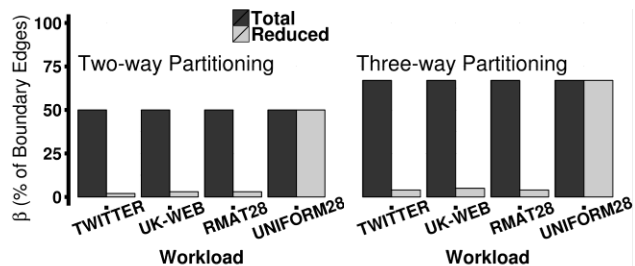

Figure 4: Resulted ratio of edges that cross partitions ($\beta$) with and without reduction for two real-world graphs (Twitter and UK-WEB), one synthetic scale-free graph (RMAT28), and one synthetic graph with uniform node degree distribution (UNIFORM28).

To highlight the benefit of reduction, we test a naïve random-based graph partitioning algorithm and compare how much communication would happen with and without reduction. Figure 4 shows $\beta$ resulted from two- and three- way partitioning, representing setups with one and two GPUs respectively, for real (Twitter and UK-WEB) and synthetic graphs (RMAT28 and UNIFORM28). The graphs are described in detail in section (§5.1); for now, the relevant characteristic that differentiates them is the degree distribution: the real-world and RMAT28 graphs are scale-free and have skewed degree distribution, while UNIFORM28 has a uniform degree distribution.

The figure shows that reduction significantly decreases $\beta$ (to less than 5%) for the graphs with skewed distribution. The worst case input is an Erdős-Renyi random graph [Erdős and Rényi 1960], which has uniform edge degree distribution. However, as discussed before, many graphs processed in practice have power-law degree distribution, thus this optimization is useful in practice.

Finally, it is important to mention that reduction works for algorithms where it is possible to reduce, at the source partition, into one value the values sent to the same remote vertex. Although some graph algorithms cannot benefit from reduction (e.g., triangle counting), we argue that a wide range of graph algorithms has this characteristic. For example, the "visited" status in BFS, minimum "distance" in SSSP, minimum "label" in a connected components algorithm, and the "rank" sum in PageRank.

### 3.5 Summary

With parameters set to values that represent realistic scenarios, the model predicts speedups for the hybrid platform, even when using naïve random partitioning. Hence, we conclude that *it is beneficial to explore this opportunity in more depth by prototyping an engine to partition graphs and process them on a hybrid platform* (described in §4). We show that the model offers good accuracy in §5, then evaluate the advantages of advanced partitioning techniques for a set of graph processing algorithms, workloads, and processing platforms (§6-§8), and compare with the performance of state of the art graph processing frameworks (§9.4)

### 4. TOTEM: A GRAPH PROCESSING ENGINE FOR HYBRID PLATFORMS

To enable application programmers to leverage hybrid platforms, we designed TOTEM – a graph processing engine for hybrid and multi-GPU single-node systems. This section presents TOTEM's programming model (§4.1 and §4.2), its implementation (§4.3), and a discussion of its design trade-offs (§4.4).

### 4.1 Programming Model

TOTEM adopts the Bulk Synchronous Parallel (BSP) computation model [Valiant 1990], where processing is divided into rounds – *supersteps* in BSP terminology. Each superstep consists of three phases executed in order: in the *computation phase*, each processing unit executes asynchronously computations based on values stored in their local memories; in the *communication phase*, the processing units exchange the messages that are necessary to update their statuses before the next computation unit starts; finally, the *synchronization phase* guarantees the delivery of the messages. Specifically, a message sent at superstep $i$ is guaranteed to be available in the local memory of the destination processing unit only at superstep $i+1$.

Adopting the BSP model allows to circumvent the fact that the GPUs are connected via the higher-latency PCI-E bus. In particular, batch communication matches well BSP, and enables TOTEM to partially hide the bus latency.

In more detail, TOTEM performs each of these phases as follows:
- *Computation phase.* TOTEM initially partitions the graph and assigns each partition to a processing unit. In each compute phase, the processing units work in parallel, each executing a user-specified kernel on the set of vertices that belongs to its assigned partition.
- *Communication phase.* TOTEM enables the partitions to communicate via boundary edges. The engine stores messages sent to remote vertices in local buffers that are transferred in the communication phase to the corresponding remote partitions. To reduce the communication volume, the source processor combines the messages targeted to the same remote destination vertex (as discussed in §3.4). Note that the *synchronization phase* is performed implicitly as part of the communication phase.
- *Termination.* The engine terminates execution when all partitions vote to finish in the same superstep. At this point, the engine invokes another user-specified callback to collect the results from all partitions.

### 4.2 A Programmer's View

A programmer prepares TOTEM to execute a graph algorithm by providing a number of callback functions that are executed at different points in the BSP execution cycle.

The TOTEM framework itself is essentially in charge of implementing the callback API and orchestrating these calls. This hides some of the inherent complexity of developing for a hybrid platform as TOTEM offers a common data representation, abstracts the communication through boundary edges, and hides various low-level optimizations that target the hybrid platform. For example, TOTEM optimizes the data layout to increase access locality, enables transparent and efficient communication between the processing elements, and provides abstractions to handle transparently boundary edges).

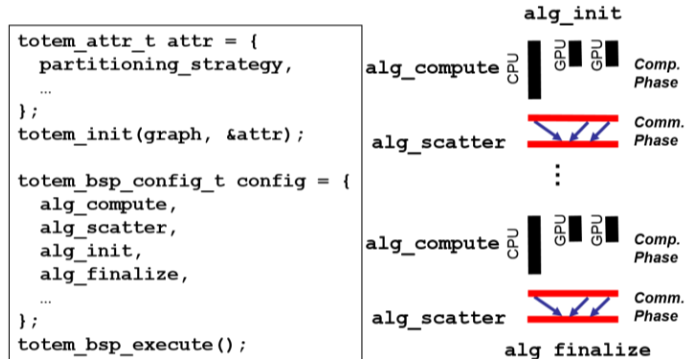

**Figure 5: A simplified TOTEM initialization presenting algorithm callbacks' mapping to BSP phases. Appendix 1 presents in detail the code for each of these callbacks for BFS.**

Figure 5 shows a simplified implementation of a graph algorithm using TOTEM (Appendix 1 presents in detail and with extensive comments how each of these callbacks looks for implementing BFS). TOTEM loads the graph and creates one partition for the host and a partition for each GPU. TOTEM accepts a number of attributes, most notably is the graph partitioning strategy (discussed in §6) and the size of each partition. The BSP engine is configured with the algorithm-specific callbacks provided by the user. The `alg_init` callback allows allocating algorithm-specific state (such as the 'level' array in BFS or the 'rank' array in PageRank), the `alg_compute` callback performs the core computation of the algorithm, while `alg_scatter` callback defines how a message received from a boundary edge updates a vertex's state (e.g., update the vertex's state with the sum of the two in the case of PageRank, or the minimum in SSSP). Finally, the `alg_finalize` callback enables the framework to release state allocated at initialization. All callbacks are invoked per partition in each BSP round.

Note that the programmer has to provide CPU and GPU versions of these callbacks. While this requires an extra effort, this gives him/her the flexibility to choose the parallel implementation that best suits each processing element. Each callback has access to the entire graph state stored on the processing element where it executes: this is a programming paradigm that has recently been dubbed "think like a graph" (as opposed to "think like a vertex") [Tian et al. 2013].

### 4.3 TOTEM Design and Implementation

TOTEM is open-source, and is implemented in C and CUDA. While a number of aspects related to TOTEM's design and implementation are worth discussing, for brevity we discuss only two: the data structures used to represent the graph and communication via boundary edges.

*4.3.1 Graph Representation and Additional Data Structures to Support Partitioning*

Graph partitions are represented as Compressed Sparse Rows (CSR) in memory [Barrett et al. 1994], a space-efficient graph representation that uses $O(|V| + |E|)$ space. Figure 6 shows an example of the CSR memory layout and the supporting data structures for a two-way partitioning setup. The arrays $V$ and $E$ represent the CSR data structure. In each partition, the vertex IDs span a linear space from zero to $|V_p|$-1. A vertex ID together with a partition ID represents a global ID of a vertex. A vertex accesses its edges by using its ID as an index in $V$ to fetch the start index of its neighbors in $E$.

The array $E$ stores the destination vertex of an edge, which includes the partition ID (shown in the figure as subscripts) encoded in the high-order bits. In the case of boundary edges, the value stored in $E$ is not the remote neighbor's ID, rather it is an index to its entry in the outbox buffer (discussed later). To simplify state management, a vertex in a directed graph has access only to its outgoing edges, which is sufficient for most graph algorithms (undirected edges can be represented as two directed edges, one in each direction).

The array $S$ represents the algorithm-specific local state for each vertex, it is of length $|V_p|$, and is indexed using vertex IDs. A similar array of length $|E_p|$ can be used if the state is required per-edge rather than per-vertex.

The processing of a vertex typically consists of iterating over its neighbors. A neighbor ID is fetched from $E$, and is used to access $S$ for local neighbors, or the outbox buffer for the remote ones. Typically, accessing the state of a neighbor (either in $S$ or in the outbox buffer) is done via atomic operations as multiple vertices may simultaneously try to update the state of a common neighbor.

To improve pre-fetching, the set of neighbors of each vertex in *E* are ordered such that the local edges are processed first (entails accessing *S*), and then the boundary edges (entails accessing the outbox buffers).

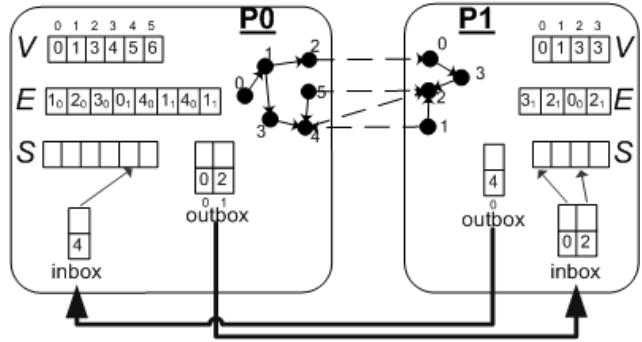

**Figure 6: An illustration of the graph data structure and the communication infrastructure in a two-way partitioning setup.**

*4.3.2    Communication via Boundary Edges*

A challenge for a graph processing engine for hybrid setups is keeping the cost of communication low. TOTEM addresses this problem by using local buffers and user-provided reduction callbacks. Messages sent via boundary edges in the computation phase of a superstep are temporarily buffered and, if possible, aggregated in these buffers then transferred in the communication phase.

TOTEM maintains two sets of buffers for each processing unit (Figure 6). The outbox buffers have an entry for each remote neighbor, while the inbox buffers have an entry for each local vertex that is remote to another partition. An in/outbox buffer is composed of two arrays: one maintains the remote vertex ID and the other stores the messages.

The outbox buffer in a partition is symmetric to an inbox buffer in another. Therefore, in the communication phase, only the message array is transferred. Once transferred, TOTEM uses the user-defined reduction function (`alg_scatter` in Figure 5) to update the remote neighbors' state in the *S* array at the remote partition with the new values. Similar to array *E*, the entries in the inbox buffers are sorted by vertex IDs to improve pre-fetching and cache efficiency when doing the update.

Note that TOTEM allows for two-way communication via the boundary edges: a vertex can either "*push*" updates to its neighbors, or "*pull*" (i.e., read) the neighbors state to update its own value. This is a necessary feature for some graph algorithms (e.g., Betweenneess Centrality) and an optimization for others (e.g., PageRank).

*4.3.3    Space Complexity of a Partitioned Graph*

In this section, we summarize the space complexity of a graph partition in TOTEM. The space cost consists of the following:

(i) The graph data structure, which has a space complexity of $O(|V_p| + |E_p|)$, where $|V_p|$ and $|E_p|$ represent the number of vertices and edges in the partition respectively. The actual size of the graph data structure $eid \times |V_p| + vid \times |E_p|$ bytes, where *eid* is 4 bytes if the graph contains less than 4 Billion edges and 8 bytes if more, while *vid* is 4 bytes if the graph contains less than 4 Billion vertices, and 8 bytes if more.

(ii) The inbox buffer, which has a space complexity of $O(|V_i|)$, where $|V_i|$ represents the number of vertices in the partition that are remote to other partitions. The actual size is $(vid+s) \times |V_i|$ bytes, where *s* is the size of the state to be communicated, which is algorithm specific (e.g., 4 bytes representing the rank in PageRank or the distance in the case of SSSP).

*(iii)* The outbox buffer, which has space complexity of $O(|V_o|)$, where $|V_o|$ represent the number of vertices in other partitions that are remote to this partition. The actual size is $(vid+s) \times |V_o|$ bytes.

In total, the space complexity of a graph partition can be expressed as $O(|V_p| + 2 \times |E_p| + |V_i| + |V_o|)$, while the actual size is equal to $eid \times |V_p| + vid \times |E_p| + w \times |E_p| + (vid+s) \times |V_i| + (vid+s) \times |V_o|$.

In addition to the above mentioned costs, each partition stores algorithm-specific state on vertices or edges. For example, BFS stores a number per vertex representing its distance, while PageRank stores a floating point number representing the rank of each vertex.

Finally, as an example, and in order to get a better sense of the memory footprint in practice, Section 9.4 includes a presentation of the actual memory footprint (specifically Table 5) of a GPU partition for the different algorithms when processing a real-world workload.

*4.3.4 Summary of Other Optimizations*

In the following we summarize the main optimizations employed by TOTEM. They have been discovered through an iterative exploration process and provide sizeable gains.

*(i)* Sorting vertex IDs in the inbox buffers to improve pre-fetching and cache efficiency when updating the vertices' local state.
*(ii)* Processing the local and remote edges separately to improve data access locality.
*(iii)* For large-scale graphs, the *V* and/or *E* arrays of GPU partitions can be allocated on the host (as mapped memory) to enable assigning larger portion of the graph to the memory-limited GPUs. Note that those arrays are immutable, and they are accessed sequentially during an algorithm execution, hence allowing for coalesced memory access reads via the high-bandwidth PCI-E bus.
*(iv)* Overlapping communication with computation to hide communication overhead. For example, if the GPU finishes processing its partition faster than the CPU does, the GPU will start copying its output buffer to the CPU's input buffer while the CPU still processing its partition, and vice versa. Double buffering techniques enable such an optimization.

**4.4 Design Trade-offs**

There are two main trade-offs in the current TOTEM implementation that are worth discussing. First, the graph representation (CSR) used makes it expensive to support updates to the graph structure during algorithm execution (e.g., creation of new edges or vertices). This is a tradeoff, as CSR enables a lower memory footprint and efficient iteration over the graph's elements (vertices and edges), which are essential for performance. Any other graph data structure that enables mutable graphs will have to have some form of dynamic memory management (e.g., linked lists), which is costly to support, particularly on GPUs.

Our decision is based on the fact that a large and important class of applications is based on static graphs. For example, many graph-based applications in social networks [Gupta et al. 2013; Wang et al. 2013] and web analytics [Malewicz et al. 2010] are performed on periodic snapshots of the system's state, which is typically maintained in storage efficient, sometimes graph-aware, indexing systems [Curtiss et al. 2013; Barroso et al. 2003].

The second limitation is related to the way communication is performed. During the communication phase of each superstep, the current implementation copies the whole outbox buffer of a partition to the inbox buffer of a remote partition assuming that there is a message to be sent via every edge between the two partitions. This is efficient for algorithms that communicate via *each* edge in every superstep, such as PageRank. However, this is an overhead for algorithms that communicate only via a selective set of edges in a superstep (e.g., in the level-synchronized BFS algorithm, at a given superstep, only the vertices in the *frontier* communicate data via their outgoing edges). Additional compression techniques can be employed to lower the communication volume.

## 5. EVALUATING MODEL ACCURACY AND PROCESSING OVERHEADS

This section aims to address the following questions: First, *how does TOTEM performance compare to that predicted by the model?* Answering this question (§5.1) allows us to validate the model and understand, for each use case, how much room is left for optimizations.

Second, we evaluate *on which phase (computation or communication) and processing element (CPU or GPU) the bulk of time is spent?* Such profiling (§5.2) identifies the bottlenecks in the system, and guides our quest for better performance.

***Testbed Characteristics.*** We use a machine with state-of-the-art (as of writing this paper) CPU and GPU models (Table 1). The two processing elements are representative for their categories and support different performance attributes. On the one hand, GPUs have a significantly larger number of hardware threads, higher memory access bandwidth, and support a larger number of in-flight memory requests. On the other hand, the CPU cores are clocked at over double the frequency, and have access to roughly one order of magnitude larger memory and cache. Finally, the GPU we use is in the same price range as its CPU counterpart.

**Table 1: Testbed characteristics: two Xeon 2560 processors and two GeForce Kepler Titan GPUs, connected via PCI-E 3.0 bus.**

| Characteristic | Sandy-Bridge (Xeon 2650) | Kepler (Titan) |
|---|---|---|
| Number of Processors | 2 | 2 |
| Cores / Proc. | 8 | 14 |
| Core frequency (MHz) | 2000 | 800 |
| Hardware Threads / Core | 2 | 192 |
| Hardware Threads / Proc. | 16 | 2688 |
| LLC / Proc. (MB) | 20 | 2 |
| Memory / Proc. (GB) | 128 | 6 |
| Mem. Bandwidth / Proc. (GB/s) | 52 | 288 |
| TDP / Proc. (Watt) | 95 | 250 |
| Price ($) | 1,171 | 960 |

***Benchmarks.*** We evaluate in detail four graph algorithms with different characteristics: Breadth-first Search (BFS), Betweeness Centrality (BC), PageRank, and SSSP. We also present a brief evaluation of Connected Components in Section 9.4. The details of the algorithms and their implementations are discussed in later sections. However, one difference between the algorithms is worth mentioning here: BFS uses a summary data structure, a bitmap, to increase the utilization of the cache, while the others do not.

***Workloads.*** We use an instance of Graph500 workload, RMAT28 graph [4] (all workloads are presented in Table 2). The memory footprint of this workload is large compared to the space available on a single GPU (~4 times larger), yet it allows us to explore offloading ratios as high as 50% when using a second GPU.

***Time Measurements.*** For all experiments in this and the following sections, we measure the time to execute the algorithm only. The time to load and partition the graph is not included when calculating the processing rate of an algorithm. Separating the algorithm processing time from the time spent on pre-processing the graph is common [Nguyen et al. 2013] as the pre-processing time is considered an amortized cost. Note that the Graph500 challenge also adopts this approach, where only the algorithm's processing time is used for ranking.

---

[4] The RMAT graphs are described by the log base 2 of the number of vertices (e.g., RMAT30 graph has $2^{30}$ vertices). Unlike in the Graph500 challenge, our graphs are directed (as generated by the model).

Table 2: Workload characteristics. The synthetic graphs were generated using the Recursive MATrix (RMAT) process [Chakrabarti et al. 2004] with the following parameters: (A,B,C) = (0.57, 0.19, 0.19) and an average vertex degree of 16.

| Workload | |V| | |E| | Memory (GB) |
|---|---|---|---|
| Twitter [Cha et al. 2010] | 52M | 1.9B | 7,689 |
| UK-Web [Boldi et al. 2008] | 105M | 3.7B | 14,666 |
| RMAT27 | 128M | 2.0B | 8,704 |
| RMAT28 | 256M | 4.0B | 17,048 |
| RMAT29 | 512M | 8.0B | 36,864 |
| RMAT30 | 1,024M | 16.0B | 73,728 |

*Evaluation Metrics: Execution Time and Traversed-Edges-per-Second (TEPS).* While in this section we report speedups when comparing with a host-only execution, later sections report TEPS as a performance metric. Similar to the Graph500 benchmark, the corresponding TEPS for BFS is calculated by dividing the sum of the degrees of the visited vertices by the time. The way we calculate TEPS for SSSP and BC is similar. For SSSP, the number of edges traversed is calculated by summing the degrees of the vertices that have a non-infinite distance; for BC, a non-zero score, with the difference being that for BC the number of traversed edges is multiplied by two as the algorithm has backward and forward propagation phases (see Section 7.2 for details regarding the BC algorithm). Finally, for PageRank, the corresponding TEPS is computed by dividing the number of edges in the graph by the time per PageRank iteration (in each iteration, each vertex accesses the state of all its neighbors). We believe the TEPS metric has the advantage that it can allow a (rough) comparison between runs of the same algorithm or implementation on different workloads.

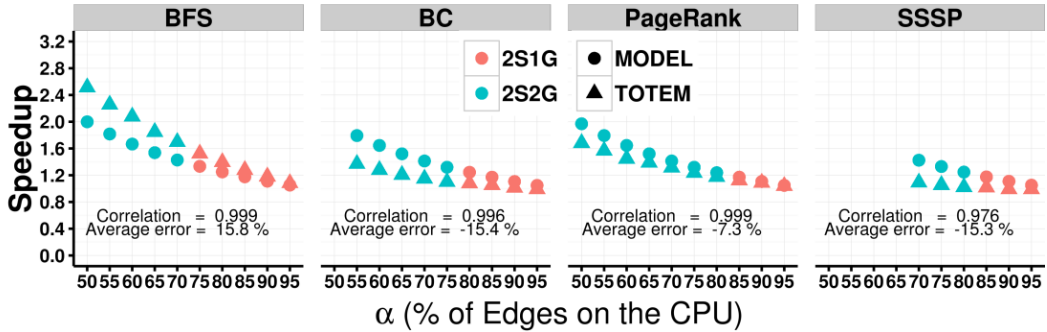

Figure 7: Speedup predicted by the model (circles) and achieved by Totem (triangles) for RMAT28 graph while varying the percentage of edges assigned to the CPU partition. The plot shows the results while offloading random partitions to one (2S1G in red) and two (2S2G in blue) GPUs. The $r_{cpu}$ used to calculate the speedup predicted by the model is the CPU-only performance on our experimental machine (the absolute TEPS are reported in Figure 23). Note that the start point on the x-axis represents the minimum percentage of edges that needs to be kept on the host due to GPU space constraints. Also, note that due to different memory space requirements between algorithms, the point at which a second GPU needs to be used is different for each algorithm. Pearson's correlation coefficient [Lee Rodgers and Nicewander 1988] is reported on each plot: a value in the range [1,-1] where 1 is total positive correlation and 0 is no correlation. Average error and correlation for other workloads are reported in Table 3.

*Data Collection and Notations.* For each data point, here and in later evaluation sections, we present the average over 64 runs. Error bars present the 95% confidence interval, in most cases, are too narrow to be visible.

The different hardware configurations used in our experiments are presented in the following notation: *xS yG*, where *x* is the number of CPU sockets (processors) used, while *y* represents the number of GPUs. For example, "*2S1G*" refers to processing the graph on two CPU sockets and one GPU.

### 5.1 Totem and the Performance Model

We first compare the speedup predicted by the model and the one achieved by TOTEM. Figure 7 shows the speedup while varying $a$, the percentage of edges left on the CPU for the four graph algorithms for RMAT28 workload. Note that the figure shows the speedup while using one (2S1G in blue) and two (2S2G in red) GPUs. Table 3 presents a summary of the correlation coefficients and average errors for all other workloads.

We observe the following: First, the achieved speedup has strong positive correlation with the one predicted by the model for all algorithms and with low average error. Second, the model underpredicts BFS performance. This is because for BFS, offloading to the GPU not only reduces the amount of work that the CPU needs to do, but also improves the CPU processing rate due to improved cache hit ratio: the bitmap used by BFS becomes smaller and hence fits better into the cache (Section 6.3.2 presents experiments that support this hypothesis). This effect is not captured by the model.

Table 3: Average error and correlation between the predicted speedup by the model and the achieved one by Totem for all algorithms and workloads.

| Algorithm | Workload | Correlation | Avg. Err. |
|---|---|---|---|
| BFS | RMAT27 | 0.99 | 6% |
|  | RMAT28 | 0.99 | 16% |
|  | RMAT29 | 0.99 | 6% |
|  | RMAT30 | 0.99 | 11% |
|  | Twitter | 0.99 | -1% |
|  | UK-WEB | 0.99 | -25% |
| PageRank | RMAT27 | 0.99 | 4% |
|  | RMAT28 | 0.99 | -7% |
|  | RMAT29 | 0.97 | 4% |
|  | RMAT30 | 0.99 | 8% |
|  | Twitter | 0.93 | 10% |
|  | UK-WEB | 0.98 | -8% |
| BC | RMAT27 | 0.99 | -13% |
|  | RMAT28 | 0.99 | -15% |
|  | RMAT29 | 0.99 | -10% |
|  | RMAT30 | 0.99 | -3% |
|  | Twitter | 0.99 | -11% |
|  | UK-WEB | 0.99 | -5% |
| SSSP | RMAT27 | 0.98 | -20% |
|  | RMAT28 | 0.97 | -15% |
|  | RMAT29 | 0.99 | -8% |
|  | Twitter | 0.88 | -22% |
|  | UK-WEB | 0.97 | -4% |

The latter observation is important as it suggests that carefully choosing the part of the graph to be offloaded to the GPU may lead to superlinear speedups due to cache effects. We evaluate this premise in more detail in Section 6.

Finally, we note that similar accuracy holds for other workloads. Moreover, we have shown in a previous work [Gharaibeh et al. 2012] this also holds for a different hardware platform. We do not present these results here for brevity.

### 5.2 Overhead Analysis

To understand on which phase (computation or communication) and processing element (CPU or GPU) the bulk of time is spent, we look at the breakdown of the total execution time. Figure 8 shows the percentage of time spent on each phase for BFS while processing RMAT28 graph. (Note that the other algorithms exhibited the exact same behavior, moreover these results were observed on all other workloads.)

Two points are worth discussing: First, the GPU processes its partition at a faster rate, as a result, processing the CPU partition always remains the main bottleneck. The GPU is 2 to 20 times faster; this indicates that our assumption that the GPU finishes its processing first holds in practice.

Second, the CPU-GPU communication overhead is significantly lower than the computation, even when using two GPUs. This is due to aggregating boundary edges and to the high bandwidth of the PCI-E bus.

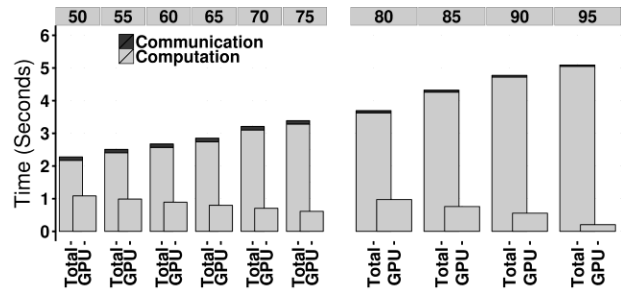

Figure 8: Breakdown of BFS execution time for the RMAT28 graph (the same data points in Figure 7). *Left*: using two GPUs (2S2G). *Right*: using one GPU (2S1G). The "Computation" bar shows the time of the bottleneck processor (the CPU in this case). The GPU partition(s), processed concurrently, is shown for comparison.

The fact that *communication is not a bottleneck* has important consequences: *rather than focusing on minimum cuts when partitioning the graph to reduce* communication (a pre-processing step that, generally, is prohibitively expensive), *an effective partitioning strategy should focus on reducing computation.*

To this end, the next section explores the impact of various graph partitioning strategies and workload allocation schemes on the performance of graph algorithms on a hybrid system. Particularly, we focus on investigating low-cost partitioning techniques that *generate workloads that match well the strength of the processing element they are allocated to*.

## 6. GRAPH PARTITIONING FOR HYBRID SYSTEMS

This section presents the set of requirements for effective partitioning strategies for hybrid systems (§6.1), discusses (§6.2) and evaluates (§6.3) our proposed degree-based partitioning strategy.

### 6.1 Partitioning Strategy Requirements

An effective graph partitioning strategy must have the following characteristics:
- *Minimizes algorithm's execution time by reducing computation (rather than communication).* The BSP model divides processing into computation and communication phases. We focus on partitioning strategies that reduce the computation time. We note that our approach is in sharp contrast to previous work on graph partitioning for distributed graph processing, as they focus on minimizing the time spent on communication (e.g., by minimizing the edge-cut between partitions) [Chamberlain 1998]. Our evaluation in the previous section (§5.2) provides the intuition that supports this choice: message reduction and batch communication (assisted by the high bandwidth of the PCI-E bus that typically connects discrete GPUs) can significantly reduce the communication overhead for concurrent graph processing (or similar applications, as the optimizations are application agnostic) on hybrid systems, which makes computation rather than communication the bottleneck.
- *Has a low space and time complexity*. Processing large-scale graphs is expensive in terms of both space and time; hence partitioning algorithms with time complexity higher than linear or quasilinear are impractical.
- *Handles large scale-free graphs*. Many important graphs in different domains present skewed vertex degree distributions. Therefore, the partitioning strategy must be able to handle the severe workload imbalance associated with such large-scale graphs.

## 6.2 Partition by Degree Centrality

We propose to partition the graph by degree centrality, placing the high-degree vertices in one type of processor and the low-degree ones in the other type. Our hypothesis is that this simple and low-cost partitioning strategy brings tangible performance benefits while meeting the solution requirements.

The motivation behind this intuition is twofold. First, dividing a scale-free graph using the vertex degree as the partition criterion produces partitions with significantly different levels of parallelism that match those of the different processing elements of the hybrid system. Second, such a partitioning strategy produces partitions that are more homogenous in terms of vertex connectivity compared to the original graph, resulting in a more balanced workload within a partition. This is important to maximize the utilization of a processor's cores, especially for the GPU because of its strict parallel computation model.

Partitioning the graph based on vertex degree is low cost in terms of computational and space complexity. One way to classify the low and high degree vertices is by sorting, with time complexity $O(|V|\log|V|)$. In practice, one can improve the running time even further by using partial sorting (i.e., finding the degree values that divide the graph into the desired partitions), which takes linear $O(|V|)$ time complexity [Chambers 1971]. Regarding space complexity, these manipulations require $O(|V|)$ of additional space, which represent the permuted vertex ids after sorting (or partial sorting). Once the vertices are placed in the required order, the edges of each vertex can be read from disk according to the new order. This is a moderate space cost as the size of scale-free graphs is typically dominated by the number of edges.

## 6.3 Evaluation

### 6.3.1 Highlighting the Effect of Partitioning

We use the BFS benchmark to evaluate the partitioning strategies. We compare three partitioning strategies: RAND, HIGH, and LOW. RAND divides the graph randomly. The other two strategies are based on degree centrality: HIGH divides the graph such that the highest degree vertices are assigned to the CPU, and LOW divides the graph such that the lowest degree vertices are assigned to the CPU.

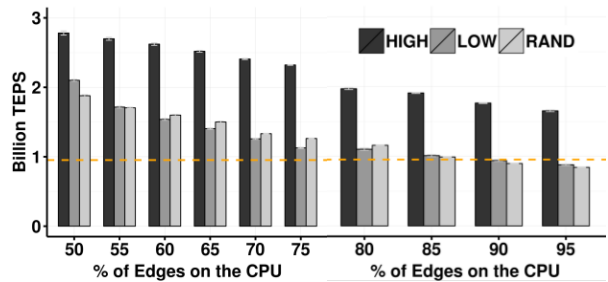

**Figure 9: BFS traversal rate (in billions of traversed edges per second - TEPS) for the RMAT28 graph and different partitioning algorithms while varying the percentage of edges placed on the CPU. *Left*: two GPUs (2S2G); *Right*: one GPU (2S1G). The performance of processing the whole graph on the host only (2S) is shown as a straight dashed line.**

Figure 9 shows BFS traversal rate in billions traversed edges per second (TEPS) for the RMAT28 workload (|V|=256M, |E|=4B). Note that the graph is too large to fit entirely on one or two GPUs and, thus, the host must keep at least 80% and 50% of the graph's edges, respectively.

In this figure, the x-axis represents the share of the edge array assigned to the CPU partition (after the vertices in the vertex-array have been ordered by degree). For example, consider the 80% data point and HIGH partitioning. The high-degree vertices are assigned to the host until 80% of the edges of the graph and their corresponding vertices are placed on the host. The remaining vertices and their edges are placed on the GPU. Similarly, in the case of LOW partitioning, the low-degree vertices are assigned to the host until it holds 80% of the graph's edges.

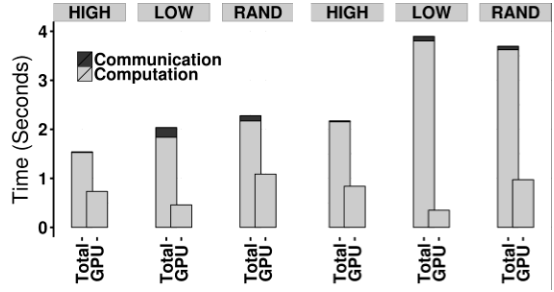

**Figure 10: Breakdown of execution time for the RMAT28 graph. *Left*: using two GPUs and 50% of the edges are assigned to the CPU. *Right*: using one GPU and 80% of the edges are assigned to the CPU. The Total bar refers to the total execution time (i.e., the makespan). The Computation portion of the Total bar refers to the time of the bottleneck processor (the CPU in all cases). The GPU bar refers to the portion of Computation time where the GPU was busy.**

The figure reveals a significant performance difference generated by the various partitioning schemes. In particular, assigning the high-degree nodes to the CPU results in superlinear speedup with respect to the share of the graph offloaded for processing on the GPU. For example, offloading 50% of the graph to be processed on the GPUs offers 2.8x speedup. The next subsection explores the causes of this speedup.

*6.3.2 Explaining the Performance Difference*

Figure 10 presents the breakdown of execution time for two of the data points presented in Figure 9: the 50% and 80% data points, which represent the maximum partition size that can be offloaded to two and one GPU(s), respectively. The breakdown shows that the hybrid system's performance is *bottlenecked by the CPU regardless of the partitioning scheme*, even when offloading 50% of the edges to be processed on the GPUs. This happens because of two reasons: (i) the GPU has a higher processing rate; and (ii) the communication overhead is negligible compared to the computation phase. Based

```
1  BFS(Partition p, int level){
2    bool done = true;
3    parallel for v in p.vertices{
4      if (v.level != level) continue;
5      for (n in v.nbrs){
6        if (!p.visited.isSet(n)){
7          if (p.visited.atomicSet(n)){
8            n.level = level + 1;
9            done = false;
10   }}}}
11   return done;
12 }
```

**Figure 11: Pseudocode of the level-synchronous BFS compute kernel. The kernel is invoked in each round for each partition. The algorithm terminates when all partitions in the same round return true.**

on these two observations, the rest of this section focuses on the effect of graph partitioning strategies on CPU performance.

Figure 11 lists the pseudo-code for the BFS kernel used in our implementation. Hong et al. [Hong et al. 2011b] showed that this implementation has a superior performance over typical queue-based approaches. In order to reduce main memory traffic, the algorithm uses a bit-vector (lines 6 and 7 in Figure 11) to mark the vertices that have been visited, thus avoiding fetching their state from main memory.

Chhugani et al. [Chhugani et al. 2012] showed that a cache-resident "visited" bit-vector is critical for BFS performance on the CPU, and that the performance significantly drops for large graphs as the bit-vector becomes larger. For the RMAT28 workload, the size of the "visited" bit-vector is 32MB (i.e., a bit array that represents the 256M vertices) and it is only a little smaller than the total amount of last level cache (LLC) on the two CPU sockets, which is 40MB.

To evaluate cache behavior, Figure 12 shows the LLC cache miss rate (left) and the percentage of main memory accesses (right) for the different partitioning schemes.

Depending on the partitioning strategy, the "visited" vector is differently distributed between the host and the accelerator. Thus, to better understand the profiling data in Figure 12, Figure 13 shows the percentage of vertices assigned to the CPU for each graph-partitioning scheme. The two figures highlight the relation between $|V_{cpu}|$ and the cache miss rate.

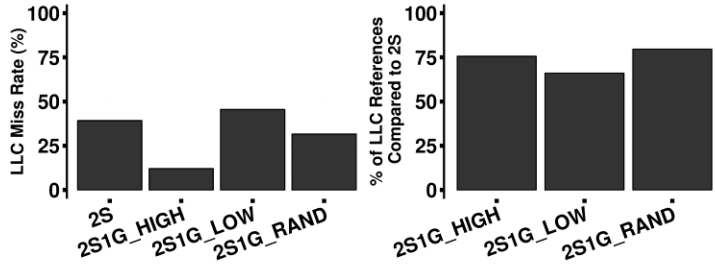

Figure 12: Performance counter statistics gathered when running BFS on an RMAT28 graph for a CPU-only configuration (2S), and a hybrid configuration using one GPU (2S1G) when 80% of the edges are assigned to the CPU. *Left*: LLC miss ratio (the lower the better), computed as $100\times(LLC\_MISS/LLC\_REFS)$. *Right*: the percentage of main memory references on the host compared to processing the whole graph on the host (the lower the better), computed as $100\times(LLC\_REFS_{2S1G}/LLC\_REFS_{2S})$.

On the one hand, RAND and LOW partitioning strategies produce a CPU partition with a large number of vertices leading to a large "visited" bit-vector similar in size to that of the original graph. Therefore, the LLC miss rate changes only slightly when compared to processing on the CPU only: improved for RAND due to lower $|V_{cpu}|$, and worsened for LOW due to the added overhead of handling boundary edges (i.e., edges with source and destination vertices reside on partitions that are assigned to different processors). However, Figure 12 (right) shows that both these strategies still reduce the number of main memory references – as a consequence of offloading part of the graph to the GPU, resulting in an overall performance improvement by the hybrid system.

On the other hand, due to the power-law degree distribution of the graph, the CPU partition produced by the HIGH strategy has two orders of magnitude fewer vertices for the same number of edges, resulting in a much more cache friendly CPU workload. This leads to a significant improvement in the CPU processing rate; as a result, the hybrid system is faster than the other two partitioning strategies.

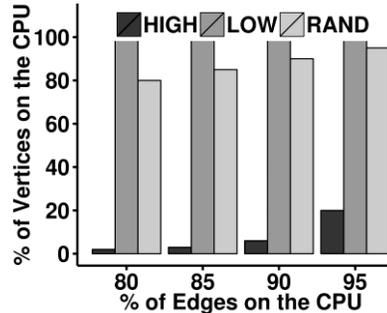

Figure 13: Percentage of vertices placed on the CPU for RMAT28 graph while varying the percentage of edges assigned to the partition, and for various partitioning strategies.

With the HIGH partitioning strategy, offloading as little as 5% of the edges to the GPU offers 2x speedup compared to processing the graph on the CPU only, and up to 2.5x speedup when offloading 25% of the edges. This demonstrates that although GPUs have limited memory, they can significantly improve the performance. Intuitively, this is because GPUs are able to efficiently handle the sparser part of the graph as they rely on massive multi-threading rather than caches to hide memory access latency.

## 7. EXTENDING THE APPLICATION SET

This section focuses on the following questions: *Do the performance gains offered by the hybrid system on BFS extend to more complex applications? How do the partitioning strategies influence performance in such settings?*

To answer these questions, we present three additional applications implemented using TOTEM: ranking web pages using PageRank (§7.1), finding the main actors in a social network using Betwenness Centrality (§7.2), and finding point-to-point shortest paths in a social network (§7.3).

### 7.1 Ranking Web Pages

PageRank [Page et al. 1999] is a fundamental algorithm used by search engines to rank web pages. In this section, we evaluate PageRank on the UK-WEB workload [Boldi et al. 2008], a crawl of over 100 million pages from the .uk domain, and 3.7 billion directed links among the pages.

```
1  PageRank(Partition p) {
2    delta =(1- dFactor)/vCount;
3    parallel for v in p.vertices {
4      sum = 0;
5      for (nbr in p.incomingNbrs) {
6        sum = sum + nbr.rank;
7      }
8      v.rank = delta + dFactor * sum;
9    }
10 }
```

**Figure 14: Pseudocode of PageRank's compute kernel.** *vCount* is the total number of vertices in the graph, while *dFactor* is the damping factor, a constant defined by the PageRank algorithm. The kernel is invoked in each BSP round for each partition. The algorithm terminates after executing the kernel a predefined number of times.

Figure 14 presents the compute kernel of the PageRank algorithm. Note that the kernel is pull-based: each vertex *pulls* the ranks of its neighbors via the incoming edges to compute a new rank. This is faster than a push-based approach, where each vertex *pushes* its rank to its neighbors via the outgoing edges. The latter approach requires atomic operations, and hence less efficient [Nguyen et al. 2013].

Compared to BFS, PageRank has a higher compute-to-memory access ratio, and does not employ summary data structures. Therefore, the cache has a lower effect on the processing performance on the host.

Figure 15 shows PageRank's processing rate. While a single GPU offers narrow improvement due to limitations on the size of the offloaded partition, adding a second GPU significantly improves the performance for such a large workload: up to 2.3x speedup compared to processing the whole graph on the CPU only.

Compared to the other two strategies, LOW partitioning allows offloading a larger portion of the edges to the GPU. This happens because PageRank requires a larger per-vertex state than BFS; hence, the number of vertices assigned to a partition has a larger effect on a partition's memory footprint. Since LOW places the high degree vertices on the GPU, the number of vertices assigned to the GPU partition by LOW is significantly lower than that assigned by HIGH and RAND strategies for the same number of edges.

Note that the HIGH strategy performs the best among all partitioning strategies. To explain this result, Figure 16 shows the breakdown of execution time. Similar to BFS, the communication overhead is negligible; the CPU is the bottleneck processor in all cases; and the HIGH partitioning is the most efficient due to faster CPU processing.

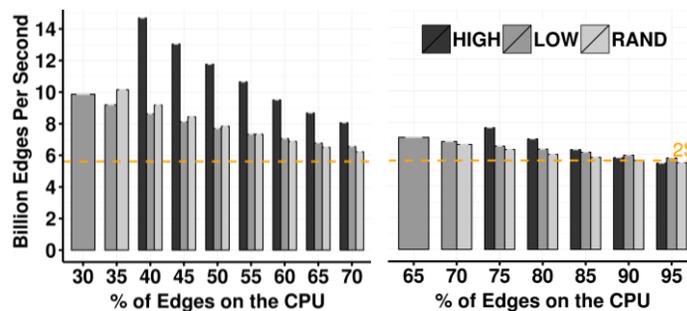

**Figure 15: PageRank traversal rate for the UK-WEB graph using one (right) and two (left) GPUs. Missing bars represent cases where the GPU's memory space is not enough to fit the GPU partitions. The performance of processing the whole graph on the two CPU sockets (2S) is shown as a line.**

Two interrelated factors lead to this result. First, from the pseudo-code in Figure 14, notice that the number of memory read operations is proportional to the number of edges in the graph (line 6), while the number of write operations is proportional to the number of vertices (line 8). Second, as discussed in the previous section, for the same number of edges, the different partitioning strategies produce partitions with drastically different number of vertices (see Figure 13). Particularly, HIGH produces a CPU partition with significantly fewer vertices. As a result, we expect that HIGH leads to a CPU partition that performs significantly fewer write operations compared to the other two strategies, while the number of read operations will be similar for all partitioning strategies.

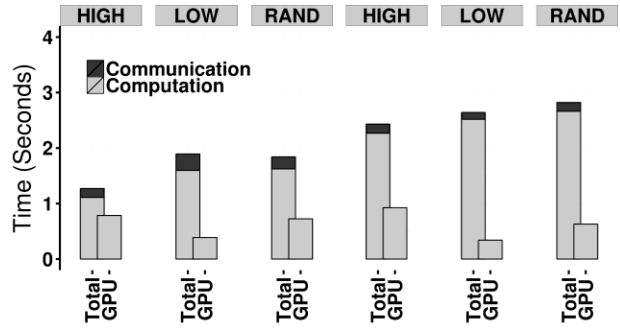

Figure 16: Breakdown of PageRank execution time (five iterations) for the UK-WEB graph when offloading the maximum size partition to two (left three bars) and one GPU (right three bars).

Figure 17 confirms this analysis: it shows the percentage of write and read memory accesses on the CPU (compared to processing the whole graph on the host) when offloading the largest possible partition to two GPUs (i.e., the percentage of edges on the CPU is 30%, 35% and 40% for LOW, RAND and HIGH, respectively). The figure demonstrates that the percentage of read accesses (Figure 17 left) is similar for all partitioning strategies, with HIGH performing slightly more reads than the other two as it allows offloading fewer edges, while the percentage of write accesses (Figure 17 right) significantly differs.

One may expect that the overhead of reads will be dominant as the number of edges is much larger than the number of vertices. However, two reasons lead to the visible impact of writes. First, the performance analysis tool LMbench [McVoy and Staelin 1996] shows that the host memory write throughput is lower, almost half, than its read throughput. Second, the reduction in the number of write accesses is significant: HIGH generates two orders of magnitude fewer write operations compared to LOW and RAND. Note that this reduction is compensated by a major increase in write memory operations in the GPU partitions, which is reflected in the increase of the GPU compute time for HIGH and RAND compared to LOW in Figure 16. Still, the GPU's high memory bandwidth allows processing this part of the workload faster than the CPU and, hence, it leads to an overall gain in performance.

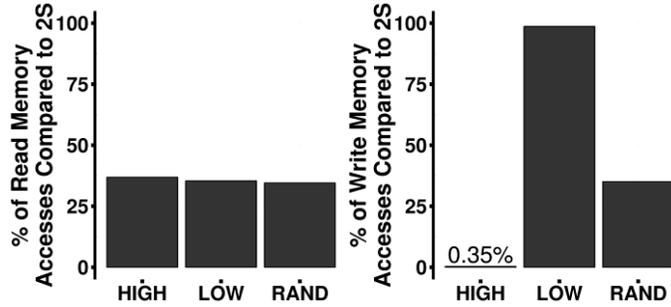

Figure 17: Host memory accesses statistics gathered when running PageRank on UK-WEB graph while when offloading the maximum size partition to two GPUs (2S2G). The performance counter used to collect these statistics is "*mem_uops_retired*". *Left*: read accesses; r*ight:* write accesses compared to processing the graph on the host only.

Finally, similar behavior is obtained for other graphs. Additionally, we show in past work [Gharaibeh et al. 2013a] that this also holds on a different machine.

### 7.2 Finding the Main Actors in a Social Network

A key measure of importance for vertices in social networks is Betweenness Centrality (BC). This section presents an evaluation of BC on a snapshot of the Twitter follower network [Cha et al. 2010]. The workload includes over 52 million users and 1.9 billion directed follower links.

We evaluate Brande's BC algorithm [Brandes 2001], which is based on forward and backward BFS traversals. Figure 18 lists the pseudocode of the forward and backward propagation kernels. Overall, the algorithm has different characteristics and is more complex than PageRank and the basic BFS algorithm presented previously. Compared to basic BFS, BC traversal does not benefit from summary data structures targeted for improving cache efficiency. Compared to PageRank, BC is a traversal-based algorithm, where the set of "active" vertices changes across iterations, and uses atomic operations.

```
1  forwardPropagation(Partition p, int level){
2   finished = true;
3   parallel for v in p.vertices{
4    if (p.dist[v] == level){
5     vNumSPs = p.numSPs[v];
6     for (nbr in v.neighbors){
7      if (p.dist[nbr] == INF){
8       p.dist[nbr] = level + 1;
9       finished = false;
10     } // if
11     if (p.dist[nbr] == level + 1){
12      atomicAdd(p.numSPs[nbr], vNumSPs);
13     } // if
14    } // for
15   }
16  }
17  return finished;
18 }

19 backwardPropagation(Partition p, int level){
20  parallel for v in  p.vertices {
21   if (p.dist[v] == level) {
22    vDelta = 0;
23    vNumSPs = p.numSPs[v];
24    for (nbr in v.neighbors) {
25     if (p.dist[nbr] == (level + 1)) {
26      vDelta += ((vNumSPs/p.numSPs[nbr])*
                    p.delta[nbr]);
27     } // if
28    } // for
29    p.delta[v] = vDelta;
30    p.betweenness[v] += vDelta;
31   } // if
32  } // for
33  return ((level – 1) == 0);
34 }
```

**Figure 18: Pseudocode of BC's compute kernels. The algorithm is executed in two BSP cycles. A first BSP cycle is run using the forward propagation kernel. Once the first cycle terminates, a second cycle is run using the backward propagation kernel.**

Figure 19 (left) shows BC processing rate while offloading part of the graph to be processed on one GPU (i.e., 2S1G configuration). The figure demonstrates that for a specific percentage of edges offloaded to the GPU, HIGH offers the best performance. Moreover, similar to PageRank, LOW partitioning allows offloading a larger percentage of the edges to the GPU than HIGH and RAND. In fact, since BC requires relatively large per-vertex state, LOW allows offloading 20% more edges to the GPU compared to HIGH. Unlike PageRank, however, offloading more edges to the GPU via LOW partitioning has a significant impact on improving the overall performance.

To understand this behavior, Figure 19 (right) shows the breakdown of overheads when offloading the maximum size partition to one GPU (i.e., the percentage of edges offloaded is 50%, 30% and 40% for HIGH, LOW and RAND, respectively). Notice that communication has minimal impact on performance, and that the CPU is again the bottleneck processor. Therefore, in the following, we quantify the major operations in the compute kernel by examining the pseudocode in Figure 18.

The major operations in the algorithm are: 5×|E| scattered reads (lines 7, 11, 12 and 26), 1×|E| atomic additions with scattered writes (line 12), 3×|E| floating point operations, 2×|V| writes (lines 29 and 30) and 1×|V| additions (line 30).

This analysis reveals that, similar to PageRank, BC performs expensive operations proportional to both the number of edges and vertices. Therefore, for a specific percentage of edges offloaded to the GPU, HIGH performs better than LOW and RAND as it results in significantly fewer vertices assigned to the bottleneck processor, the CPU. However, unlike PageRank, BC performs larger and more expensive operations per edge than per vertex. Therefore, the ability of LOW partitioning scheme to offload more edges to the GPU results in notably better performance than HIGH and RAND partitioning schemes.

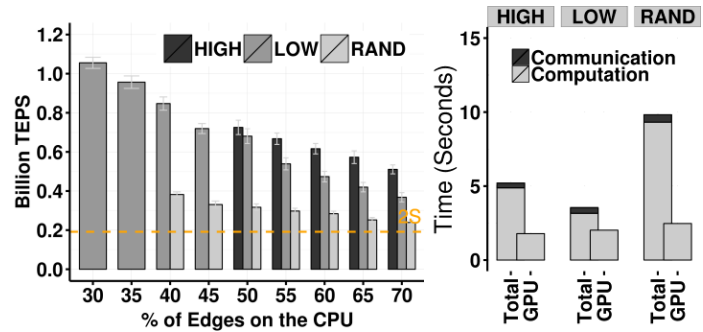

Figure 19: BC performance on the Twitter network for the 2S1G system. Left: traversal rate (in Billion TEPS) using one GPU. The horizontal line indicates the CPU-only performance using the two sockets (2S). Right: Breakdown of execution time when offloading the maximum size partition to one GPU (i.e., the percentage of edges offloaded is 50%, 30% and 40% for HIGH, LOW and RAND, respectively).

Finally, we turn our attention to comparing the performance of the hybrid system (2S1G) with the CPU only (2S) performance (the dotted line in Figure 19 (left)). Adding a GPU boosts the performance by 5x compared to the dual socket (2S) configuration, a significant improvement delivered by the hybrid platform.

### 7.3 Finding Point-to-Point Shortest Paths in a Network

The Single Source Shortest Path (SSSP) aims to find the shortest path from a given source to all connected vertices in a network. SSSP algorithms are used in a wide spectrum of application domains such as network routing, VLSI design, transportation network modeling and social network analysis. In this section, we present an evaluation of SSSP on the Twitter workload (Table 2).

Shortest path computation involves weighted graphs: each edge is associated with a piece of information commonly known as its *weight*. For example, in the Twitter follower network, where vertices represent users, an edge weight can be a measure of common followers between two users or their geographic

```
1  SSSP(Partition p) {
2    finished = true;
3    parallel for v in p.vertices {
4      if (p.active[v] == false) { continue; }
5      p.active[v] = false;
6      for (nbr in v.neighbours) {
7        new = p.dist[v] + v.weights[nbr];
8        old = p.dist[nbr];
9        if (new < old) {
10         if (old == atomicMin(p.dist[nbr], new)) {
11           p.active[nbr] = true;
12           finished = false;
13         }
14       }
15     } //for
16   } //for
17   return finished;
18 }
```

Figure 20: Pseudocode of SSSP's compute kernel based on Bellman-Ford algorithm. The array *dist* contains the computed distances of all the vertices in the partition. Each entry in the array *active* indicates the current state of a vertex. Every time a vertex's distance is updated, it becomes "active" and it may traverse its edge list in the same or the next BSP round. The algorithm terminates when there is no active vertex left. Note that atomicMin atomically updates a memory location with the new value if it is less than the current one, and returns the old value.

proximity. Weighted graphs increase memory footprint, which poses a challenge with respect to offloading a larger fraction of the graph to the GPU. The additional memory required is proportional with the number of edges.

There are three primary approaches to compute SSSP: Dijkstra's algorithm [Dijkstra 1959], Bellman-Ford algorithm [Ford 1956; Bellman 1958] and

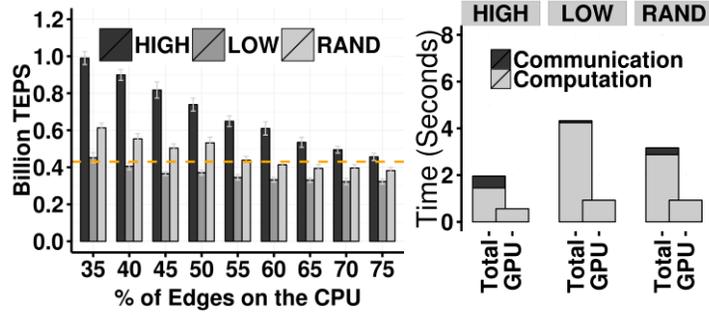

Figure 21: SSSP performance on the Twitter network for 2S2G system. Left: traversal rate (in Billion TEPS) using two GPU. The horizontal line indicates the performance of a two socket system. Right: breakdown of execution time of the 35% data point.

Delta ($\Delta$)-Stepping [Meyer and Sanders 2003]. Dijkstra's algorithm exposes no parallelism as in each round, a single vertex is made active and its corresponding edges are relaxed. $\Delta$-Stepping on the other hand groups vertices into buckets depending on their tentative distances to the source and process all the vertices within a bucket in parallel. $\Delta$-Stepping's bucket implementation, however, requires dynamic arrays which is not a good fit for GPU programming model. Furthermore, during processing, because its tentative distance may change, a vertex may need to be moved between buckets and its bucket index has to be updated. These steps require atomic operations and concurrency is hindered in the process [Davidson et al. 2014]. Bellman-Ford algorithm does not impose any constraint on the number of active vertices. In each iteration, all the vertices can relax their respective edges in parallel, making it a good fit for GPUs.

Figure 20 lists the Bellman-Ford algorithm. This is a traversal-based algorithm, but unlike BFS, the set of "active" vertices changes during an iteration and it also uses atomic operations for consistency. Our implementation of SSSP is based on the Bellman-Ford algorithm and similar to the one proposed by Harish et al. [Harish et al. 2007]. One improvement we have made is reducing the number of iterations (BSP rounds) by allowing a vertex to be set to "active" and perform "relax" operations in the same iteration if it has not been processed yet.

Figure 21 (left) shows the performance of the SSSP algorithm. As shown, HIGH partitioning offer superior performance compared to the other two portioning strategies. Figure 21 (right) shows the breakdown of execution time. Similar to BFS, PageRank and BC, communication overhead is negligible compared to that of computation. CPU is always the bottleneck processing element and the best CPU performance is achieved when HIGH partitioning is used.

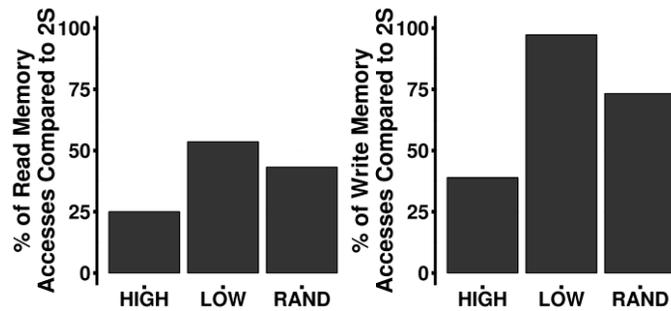

Figure 22: Host memory access statistics when running SSSP on the Twitter workload (2S2G configuration). The y-axis presents the percentage of host memory accesses of the CPU partition in a hybrid configuration compared to the number of accesses performed when running the whole graph on CPU only (i.e., 100*MEM_READ$_{2S2G}$/MEM_READ$_{2S}$ for the left figure and 100*MEM_WRITE$_{2S2G}$/MEM_WRITE$_{2S}$ for the right figure). The x-axis presents the three partitioning algorithms while offloading the maximum size partition to two GPUs.

The most critical operation in the SSSP algorithm is when a vertex atomically updates the distance of a neighbor (lines 10 to 12 in Figure 20). Having a significantly lower number of vertices in the CPU partition as a result of using a HIGH partitioning strategy contributes to reducing the contention on the atomic updates, and hence improving the overall performance of the CPU partition.

To better illustrate this analysis, Figure 22 shows host memory access statistics of the three partitioning strategies. The figure demonstrates that while all strategies lead to reduction in read memory accesses, the HIGH partitioning strategy results in a significant reduction in the number of write operations (which, as we discussed before, more expensive than read operations).

## 8. EVALUATING SCALABILITY USING SYNTHETIC WORKLOADS

This section focuses on the following questions: *How does the hybrid system scale when increasing the graph size and with various hardware configurations? What is more beneficial, adding more CPUs or GPUs?*

Figure 23 presents BFS, PageRank, BC and SSSP traversal rate for different hardware configurations (up to two sockets and two GPUs) and graph sizes (1 to 16 billion edges).

First, we focus on the analysis of configurations with two processing units. The figures show that, for all algorithms, the hybrid system (1S1G) performs better than the dual-socket system (2S). On the one hand, adding a second socket doubles the amount of last level cache and the number of memory channels, which are critical resources for graph processing performance, hence leading to close to double the performance compared to 1S configuration. On the other hand, the performance gain of 1S1G, brought by matching the heterogeneous graph workload with the hybrid system, outperforms that of the dual-socket symmetric system: between 30% to 60% improvement compared to the dual socket system (2S).

Second, the figure also demonstrates the ability of the hybrid system to harness extra processing elements. For example, in the case of BFS, the system achieves up to 3 Billion TEPS for the smallest graph ($|E|=2B$), and, more important, it achieves as high as 1.68 Billion TEPS for an RMAT30 graph ($|E|=16B$). It is worth pointing out that such performance is competitive with the performance results of the latest Graph500[5] competition for graphs of the same size. Also note that TOTEM is a generic graph-processing engine, as opposed to the dedicated BFS implementations for most submissions in Graph500; moreover the BFS implementation evaluated here is the standard top-down algorithm compared with the direction-optimized implementations [Beamer et al. 2013] that top the Graph500 competition.

Finally, the figures also demonstrate that the GPU can provide significant improvements for the large graphs, RMAT29 and RMAT30. This is made possible by employing *mapped memory* to increase the size of the offloaded partition. Particularly, for such large graphs, the GPU's limited memory space significantly constrains the size of the offloaded partition. For example, the GPUs on our testbed support 6GB of memory, and can host at most 0.625 Billion edges considering 64-bit edge identifiers (not including the space needed for the vertices' state, hence this limit is even lower especially for PageRank, BC and SSSP); therefore, the GPU's device memory can store less than 5% of graph's edges. To enable offloading a larger partition to the GPU, we allocate part of the state on host memory and map it into the GPU's address space. The tradeoff is extra communication overhead over the high latency PCI-E bus.

---

[5] www.graph500.org

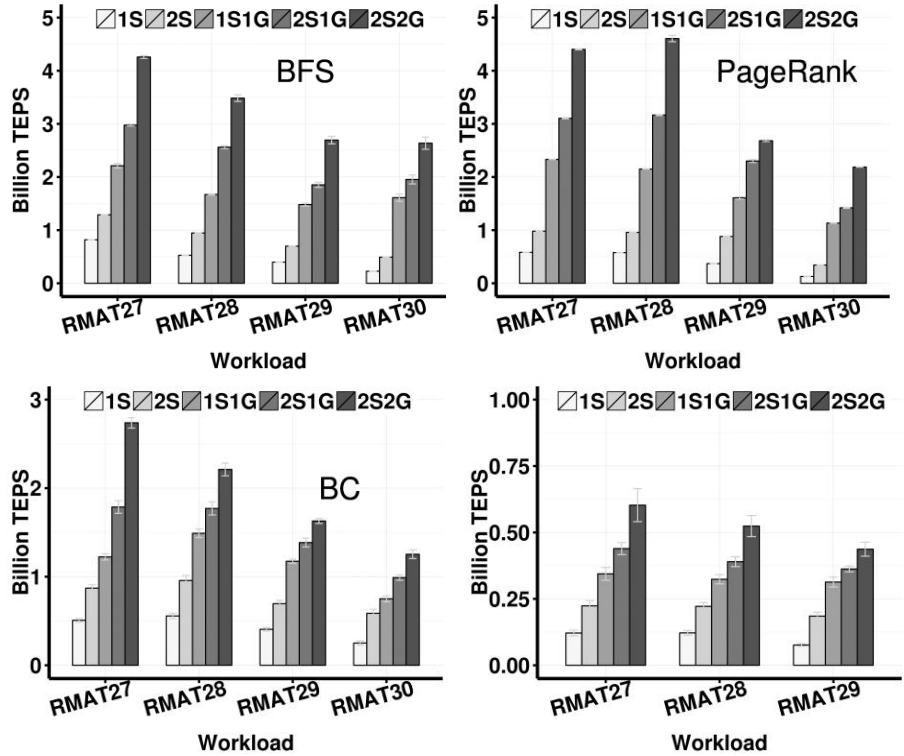

Figure 23: BFS, PageRank, BC and SSSP processing rates for different hardware configurations and R-MAT graph sizes. When GPUs are used, the graph is partitioned to obtain best performance. Experiments on configurations with a single socket (i.e., *1S* and *1S1G*) were performed by binding the CPU threads to the cores of a single socket. The result for an RMAT30 graph is missing for SSSP because of memory space constraints (SSSP requires additional memory space to store the edge-weights).

We reduce this overhead by taking the following measures: First, we reduce the impact of the high latency of the bus by restricting the use of mapped memory to allocate the part of the state that is (i) read-only, and (ii) can be accessed sequentially in batches; particularly, we used *mapped memory* to allocate the edges array since we assume static graphs. Second, we maximize transfer throughput by ensuring that the edges of a vertex are read in a coalesced manner when the vertex iterates over its neighbors.

In summary, mapped memory affects performance in the following way: for small scale graphs (RMAT28 and below), the benefit from offloading a larger partition to the GPU via mapped memory is masked by the extra overhead of reading the graph data structure via the high latency PCI-E bus (even though mapped memory by design overlaps communication and computation). For large-scale graphs (RMAT29 and above) using mapped memory was beneficial. We summarize these points in a recent poster publication [Sallinen et al. 2014].

## 9. RELATED WORK AND PERFORMANCE COMPARISON WITH OTHER SYSTEMS

This section discusses related work from several aspects: §9.1 reviews efforts on optimizing graph algorithms for multi- and many-core platforms; §9.2 reviews work related to graph partitioning; §9.3 reviews abstractions similar to TOTEM that aim to hide the complexity of implementing graph algorithms on parallel platforms; finally, §9.4 compares the performance of TOTEM with the best report numbers in the literature.

### 9.1 Optimizing Graph Algorithms

While we are unaware of previous works on optimizing graph processing on hybrid systems, many efforts exist on optimizing graph algorithms on homogeneous systems: either on multicore CPUs or on GPUs alone. For example, several studies focus on optimizing BFS on multi-core CPUs [Agarwal et al. 2010; Hong et al. 2011b; Chhugani et al. 2012]. For example, Chhugani et al. [Chhugani et al. 2012] apply a set of sophisticated techniques to improve the cache hit rate of the "visited" bit-vector, reduce inter-socket communication, and eliminate the overhead of atomic operations by using probabilistic bitmaps. Our approach to partition the graph goes in the same direction in terms of improving the cache hit rate on the CPU using a hybrid system.

Past projects have also explored GPU-only graph processing. These projects either assume that the graph fits the memory of one [Hong et al. 2011a; Katz and Kider Jr 2008], or multiple GPUs [Merrill et al. 2012]. In both cases, due to the limited memory space available, the scale of the graphs that can be processed is significantly smaller than the graphs presented in this paper.

Hong et al. [Hong et al. 2011b] work is, perhaps, the closest in spirit to our work as it attempts to harness platform heterogeneity: the authors propose to divide BFS processing into a first phase done on the CPU (as, at the beginning, only limited parallelism is available), and a second phase on the GPU once enough parallelism is exposed, having the whole graph transferred to the GPU to accelerate processing. However, this technique still assumes that the whole graph fits the GPU memory; moreover, the work is focused on BFS only.

In summary, techniques that aim to optimize graph processing for either the CPU or the GPU are complementary to our approach in that they can be applied to the compute kernels to improve the overall performance of the hybrid system. In fact, we use some of these techniques in our hybrid implementations, such as using pull-based approach in PageRank and optimizing thread allocation on the GPU [Li and Becchi 2013; Hong et al. 2011a].

### 9.2 Graph Partitioning

There is no shortage of work on graph partitioning for parallel processing. Traditionally, the problem is defined as to partition a graph in a balanced way, while minimizing the edge cut. It has been shown that this problem is NP-hard [Garey et al. 1974], therefore several heuristics were proposed to provide approximate solutions. Some heuristics, such as Kernighan–Lin [Kernighan 1970], have quadratic $O(n^2 \log n)$ time complexity, which is prohibitively expensive for the scale of the graphs we target. Multilevel partitioning techniques, such as METIS by Karypis et al. [Karypis and Kumar 1998], offer an attractive time complexity.

We believe that classical solutions do not properly address the requirements for graph partitioning on hybrid platforms. Such techniques are mainly optimized to minimize communication, which is not the bottleneck in our case. Moreover, they target homogeneous parallel platforms as they focus on producing balanced partitions, which is not sufficient for a hybrid system that has processing units with largely different characteristics.

### 9.3 Graph Processing Frameworks

A number of frameworks have been proposed to simplify the task of implementing graph algorithms at scale, which can be divided into two categories depending on the target platform. On the one hand, frameworks for shared-nothing clusters, such as Pregel [Malewicz et al. 2010] and PowerGraph [Gonzalez et al. 2012], partition the graph across the cluster nodes, and provide abstractions to implement algorithm algorithms as vertex programs run in parallel. Cluster-based solutions offer the flexibility to scale with the size of the workload by adding more nodes. However, this flexibility comes at a performance and complexity costs. Particularly, performance suffers from the high cross-node communication overhead: over one order of magnitude slower compared to single-node systems [Nguyen et al. 2013]. Moreover, the fact that the system is distributed introduces new problems such as network partition, partial failures, high latency and jitter, which must be addressed when designing the framework and when implementing algorithms on top of it, hence greatly increasing the complexity of the solution.

On the other hand, single-node platforms are becoming increasingly popular for large-scale graph processing. As discussed in §1, recent advances in memory technology make it feasible to assemble single-node platforms with significant memory space that is enough to load and process large-scale graphs for a variety of applications. Such platforms are more efficient in terms of both performance and energy, and potentially less complex to program compared to shared-nothing clusters. Examples of frameworks that capitalize on this opportunity include Ligra [Shun and Blelloch 2013], Galois [Nguyen et al. 2013] and STINGER [Ediger et al. 2012]. However, we are not aware of any frameworks that harness GPUs in a hybrid setup for large-scale graph processing.

### 9.4 Comparing Totem's Performance with Other Frameworks

Nguyen et al. [Nguyen et al. 2013] proposed a lightweight graph processing framework for single-node shared memory systems named Galois. The work compared Galois with a number of other graph processing frameworks (including Ligra, GraphLab and PowerGraph) on a quad-socket system, and demonstrated that Galois compares favorably. The largest workload that Nguyen et al. used was the Twitter network described in Table 2.

In this section, we compare TOTEM's performance with that of Galois, Table 4 shows the performance of Galois when executed on our evaluation machine (labeled 2S-Galois), and the best performance reported by Nguyen et al. in their paper (labeled 4S-Galois in the table to indicate a quad-socket configuration). The table compares the four algorithms detailed in this work as well as a fifth one, namely connected components, when processing the same Twitter graph.

**Table 4: Processing times in seconds for different algorithms and hardware configurations for the Twitter workload. The 2S-Galois column reports the performance of Galois on our evaluation machine. The performance of the four socket platform (labeled 4S-Galois) is the best performance reported by [Nguyen et al. 2013] when processing the same workload for various frameworks that include Galois, Ligra, and PowerGraph. The characteristics of the 4S platform are: Four Intel E7-4860 processors, each with 10 cores (20 hardware threads) @ 2.27GHz and 24MB of LLC per processor, hence a total of 80 hardware threads and 96MB of LLC – significantly better than our platform. Note that the processing time for PageRank is for a single round, while for BC it is for a single source.**

| Algorithm/Configuration | 2S Galois | 2S TOTEM | 4S Galois | 1S1G TOTEM | 2S1G TOTEM | 2S2G TOTEM |
|---|---|---|---|---|---|---|
| **BFS** | 5.0 | 4.0 | 2.3 | 1.1 | 0.85 | 0.4 |
| **PageRank** | 24.3 | 8.1 | 10.7 | 1.5 | 1.12 | 0.5 |
| **BC** | 29.7 | 20.8 | 12.0 | 4.8 | 3.7 | 2.5 |
| **SSSP** | 13.2 | 4.6 | 8.6 | 3.3 | 3.1 | 1.9 |
| **Connected Components** | 41.1 | 42.0 | 31.9 | 38.7 | 25.8 | 13.5 |

**Table 5: For each of the 2S2G data points in Table 4, this table shows the number of vertices and directed edges allocated to a GPU partition, its memory footprint required for graph representation (including problem data, e.g., the level for BFS), inboxes, outboxes, and algorithm state. Note that for SSSP the partition size includes the edge weights, while for Connected Components (CC) the number of edges is multiplied by two to represent undirected edges as the algorithm operates on undirected graphs.**

| Algorithm | |V| | |E| | Graph Partition Representation | Inboxes | Outboxes | Algorithm State | Total |
|---|---|---|---|---|---|---|---|
| **BFS** | 26M | 883M | 3,639MB | 404MB | 171MB | 104MB | 4,318MB |
| **PageRank** | 26M | 883M | 3,639MB | 404MB | 171MB | 208MB | 4,422MB |
| **BC** | 26M | 883M | 3,639MB | 404MB | 171MB | 416MB | 4,630MB |
| **SSSP** | 26M | 637M | 5,213MB | 617MB | 303MB | 104MB | 5,911MB |
| **CC** | 17M | 1,278M | 5,180MB | 455MB | 227MB | 104MB | 6,014MB |

First, the table demonstrates that TOTEM's performance on a 2S configuration is not only better than that of Galois on our evaluation machine, but also competitive with the best reported numbers on the 4S one, even surpassing it in the cases of PageRank and SSSP. This increases our confidence that our speedup results throughout this paper use a meaningful baseline.

Second, the hybrid configurations offer significant speedups compared to all symmetric systems (2S and 4S). In the case of BFS, while the 4S system delivers 60% better performance than 2S, a modest 1S1G hybrid configuration speeds up the performance by 3.5x compared to 2S, and 2.1x compared to 4S. Moreover, the hybrid configuration 2S2G offers over 5.5x speedup compared to 4S, the symmetric system with the same number of processing elements. The figure shows that similar significant performance improvements for the other algorithms.

Finally, while Section 4.3.3 discusses the asymptotic space complexity of a GPU partition in TOTEM, Table 5 shows the breakdown of actual memory footprint for a GPU partition. This helps to clarify where most of the space overhead comes from. Notice that the graph data structure occupies over half of the required space (and most of the space in the case of SSSP because of edge weights). The Inbox/Outbox buffers, which are required to handle communication via boundary edges, occupy almost 25% of the required space. Note that in order to enable overlapping communication with computation, the inboxes/outboxes apply double buffering, which increased the space overhead of the Inboxes/Outboxes. Lastly, the algorithm-specific state consumes less than 10%.

## 10. LESSONS AND DISCUSSION

The results presented in this work allow us to put forward a number of guidelines on the opportunity and the supporting techniques required to harness hybrid systems for graph processing problems. We phrase these guidelines as answers to a number of questions.

- *Q: Is it beneficial to use a hybrid system to accelerate large-scale graph processing?*
  *A:* Yes. One concern when considering using a hybrid system is the limited GPU memory that may render using a GPU ineffective when processing large graphs. We show, however, that it is possible to offload a relatively small portion of the graph to the GPU and obtain benefits that are higher than the proportion of the graph offloaded for GPU processing. This is made possible by exploiting the heterogeneity

A. Gharaibeh et al.of the graph workload and the characteristics of the hybrid system to reshape the workload to execute faster on the bottleneck processor.

- **Q:** *Is it possible to design a graph processing engine that is both generic and efficient?*
  **A:** Yes. A range of graph algorithms can be implemented on top of TOTEM, which exposes similar BSP-based computational model and functionality to that offered by a number of other widely accepted generic graph processing engines designed for cluster environments (e.g., Pregel). Our experiments show that being generic – that is, being able to support multiple algorithms and not only the popular Graph500 BFS benchmark – did not hinder TOTEM's ability to efficiently harness hybrid systems, and scale when increasing the number of processing elements.

    We have also implemented on top of Totem the direction-optimized BFS algorithm [Beamer et al. 2013]. The results support the main takeaways we present here. Based on this implementation, Totem's performance on a hybrid system with dual-socket and dual-GPU is capable of 10.31 Billion breadth-first search traversed edges per second on a graph with 1 Billion vertices and 16 Billion undirected edges. We have submitted this result to the Green Graph500[6] competition, and ranked 6th in the 'Big Data' category.

- **Q:** *Is the partitioning strategy key for achieving high performance?*
  **A:** Yes. The low-cost partitioning strategies we explore – informed by vertex connectivity – provide in all cases better performance than blind, random partitioning.

- **Q:** *Which partitioning strategies work best?*
  **A:** The answer is nuanced and the choice of the best partitioning strategy depends on the graph size and on the specific characteristics of the algorithm (particularly on how much state is maintained and on the read/write characteristics). If the graph is large, then the CPU will likely be the bottleneck as it is assigned the larger portion of the graph, while only a small fraction can be offloaded to the GPU. Thus, the goal of partitioning is to improve the CPU performance by producing and assigning to it the friendliest workload to its architecture. Our evaluation shows that placing the high degree vertices on the CPU offers the best overall performance: it improves the cache hit rate for algorithms that use summary data structures, and, for the ones that do not use them, it offloads most of the expensive per-vertex work to the accelerator. However, for algorithms with large state per vertex, placing the few high degree nodes on the GPU allows for offloading significantly more edges (20% more in the case of Betweenness Centrality when processing the Twitter network in §7.2), and hence better balances the load between the CPU and the GPU.

- **Q:** *Should one search for partitioning strategies that lead to higher performance by searching for partitioning solutions that reduce the communication overheads?*
  **A:** No. We show that, in the case of scale-free graphs, the communication overhead can be significantly reduced – to the point that it becomes negligible relative to the processing time – by simple reduction techniques. Reduction works well for four reasons. First, many real-world graphs have skewed connectivity distribution. Second, the number of partitions the graph is split into is relatively low (only two for a hybrid system with one GPU). Third, reduction can be applied to many practical graph algorithms, such as BFS, PageRank, Single-source Shortest Path, Betweenness Centrality and Connected Components to mention only a few. Fourth, there is practically no visible cost for reduction: conceptually, reduction moves the computation to where the data is, which must happen anyway. In contrast, partitioning algorithms that aim to reduce communication have typically high computational or space complexity and may be themselves 'harder' than the graph processing required [Feldmann 2012].

---

[6] green.graph500.org

- ***Q:*** *Is there an energy cost to the time-to-solution gains provided by using GPUs?*
  ***A:*** No. One concern is that the GPU's high peak power consumption may make an accelerated solution inefficient in terms of energy. Our experience [Gharaibeh et al. 2013b] rejects this concern: GPU-acceleration allows a faster 'race-to-idle', enabling energy savings that are sizeable for newer GPU models which are power-efficient in idle state (as low as 25W [NVIDIA 2013]). Additionally, as demonstrated in the various profiling figures in this paper (Figure 8, Figure 10, Figure 16, and Figure 19), the GPU finishes much faster than the CPU, and that allows it to go to the idle state even sooner. In a past work [Gharaibeh et al. 2013b], we present a detailed discussion and evaluation of the power and energy aspects of graph processing on hybrid systems, and we show that a hybrid system is not only efficient in terms of time-to-solution, but also in terms of energy and energy-delay product.

**APPENDIX 1**

This appendix details a simplified implementation of the BFS algorithm using TOTEM. The code is thoroughly commented, and hence relatively long. The best way to read the code is to start from the **main** function, which can be found at the end of the appendix.

```
// A structure that encapsulates per-partition algorithm-specific state.
typedef struct {
  level_t*   levels;     // One slot per vertex in the partition.
  bool*      finished;   // Refers to Totem's finish flag.
  level_t    cur_level;  // The current level being processed by the partition.
} bfs_local_state_t;

// A structure that encapsulates algorithm-specific global state, which is shared
// between all partitions.
typedef struct {
  level_t* levels;  // The final output buffer.
  vid_t    source;  // The source vertex id.
} bfs_global_state_t;
static bfs_global_state_t state_g = {0};

// A helper function that is used by the CPU and GPU compute functions to process a
// vertex. The function iterates over the vertex's neighbors, and sets their level
// if it has not been set before. The function returns false when at least one
// neighbor has been updated indicating that processing has not finished yet, which
// is eventually translated to an additional BSP round. The function returns true
// when no neighbors have been updated, which translates to termination in case the
// function returns true for all processed vertices.
static __device__ __host__
bool bfs_process_vertex(partition_t* par, bfs_state_t* state, vid_t v) {
  bool finished = true;
  if (v >= par->subgraph.vertex_count ||
      state->levels[v] != state->cur_level) { return finished; }
  for (eid_t i = par->subgraph.vertices[v];
       i < par->subgraph.vertices[v + 1]; i++) {
    const vid_t nbr = par->subgraph.edges[i];
    // The following Totem function returns a reference to the state of the neighbor.
    // If the neighbor is in the same partition, the function returns a reference
    // to the neighbor's state in the local "state->levels" array. If the neighbor
    // is remote, the function returns a reference to its state in the outbox buffer.
    level_t* nbr_level = totem_engine_get_dst_ptr(par, nbr, state->levels);

    // Update the neighbor's level if it has not been set before. Note that reduction
    // for remote neighbors happens implicitly here: all vertices in this partition
    // that has an edge to this remote neighbor would test and update the same state
    // which exist as part of the outbox buffer. During the communication phase, a
    // single value will be communicated to the partition that owns the neighbor.
    if (*nbr_level == INF_LEVEL) {
      finished = false;
      *nbr_level = state->cur_level + 1;
    }
  }
  return finished;
}

// The CPU compute kernel which is called by the compute callback if the partition
// is CPU resident.
static void bfs_compute_cpu(partition_t* par, bfs_state_t* state) {
  const graph_t* subgraph = &par->subgraph;
  bool finished = true;
  #pragma omp parallel for schedule(runtime) reduction(& : finished)
  for (vid_t v = 0; v < subgraph->vertex_count; v++) {
    finished &= process_vertex(par, state, v);
  }
  if (!finished) { *(state->finished) = false; }
}

// The GPU compute kernel, which is called by the compute callback if the partition
```

Efficient Large-Scale Graph Processing on Hybrid CPU and GPU Systems

```
  // is CPU resident.
  static __global__ void bfs_gpu_kernel(partition_t par, bfs_state_t state) {
    const vid_t v = THREAD_GLOBAL_INDEX;
    if (!process_vertex(&par, &state, v)) {
      // state.finished is a reference to a flag that is shared between all partitions.
      // Totem sets this flag to true at the beginning of each superstep. A partition
      // sets this flag to false if there are active vertices that needs to be
      // processed in the next round. Totem will launch another BSP round if any
      // partition sets this flag to false.
      *(state.finished) = false;
    }
  }

  // A wrapper for the GPU compute kernel, it configures and launches the CUDA kernel.
  static void bfs_compute_gpu(partition_t* par, bfs_local_state_t* state) {
    dim3 blocks, threads;
    totem_kernel_configure(par->subgraph.vertex_count, &blocks, &threads);
    bfs_gpu_kernel<<<blocks, threads, 0, par->stream>>>(*par, *state);
  }

  // The compute callback function. Totem calls this function for each partition as
  // part of the BSP compute phase. Depending on the partition's processor, this
  // function calls either the CPU or the GPU kernel.
  static void bfs_compute(partition_t* par) {
    bfs_local_state_t* state = (bfs_local_state_t*)par->algo_state;
    if (par->processor.type == PROCESSOR_CPU) {
      compute_cpu(par, state);
    } else if (par->processor.type == PROCESSOR_GPU) {
      compute_gpu(par, state);
    }
    state->cur_level++;
  }

  // The callback to "scatter" the messages received from remote partitions to the
  // partition's local state. Totem invokes this callback at the end of the
  // communication phase after the data has been copied from the outbox buffers of
  // the remote partitions to the inbox buffers of this partition.
  static void bfs_scatter(partition_t* par) {
    bfs_local_state_t* state = (bfs_local_state_t*)par->algo_state;
    // For each message in the inbox buffer, the following template function computes
    // the minimum of the value in the message and the one the vertex currently
    // have in the local state (i.e., state->levels). The minimum is then stored in
    // the local state as the vertex's new level.
    totem_engine_scatter_inbox_min(par->id, state->levels);
  }

  // Callback to collect the final result from the partitions' local "levels" array
  // to the final output array that will be returned to the user.
  static void bfs_collect(partition_t* par) {
    bfs_local_state_t* state    = (bfs_local_state_t*)par->algo_state;
    // The following Totem function copies each value in the local state->levels array
    // to its corresponding entry in the final state_g.levels array. To do this, the
    // function uses a "map" that maps each vertex in the partition from its local
    // id space (the vertex id within the partition which is used to index the local
    // "state->levels" array) to its global id space (the vertex id in the original
    // graph which is used to index the final output array "state_g.levels").
    totem_engine_collect(par->id, state->levels, state_g->levels);
  }

  // Callback to allocate and initialize a "bfs_local_state_t" structure, a per-
  // partition and algorithm-specific state. This is called for each partition by
  // Totem at the beginning before the first BSP superstep.
  static void bfs_init(partition_t* par) {
    // Removed for brevity. In summary, the function allocates a bfs_local_state_t
    // structure for this partition. "par->alg_state" is the reference to the allocated
    // structure. It also initializes the allocated local state, such as setting
    // the level of the source vertex to 0 (if it belongs to this partition).
  }
```

```c
// Callback to free the buffers allocated in initialize. This is called by Totem at
// the end (i.e., after all partitions vote for termination).
static void bfs_finalize(partition_t* par) { // Removed for brevity. }

// The hybrid BFS algorithm entry function. Given a graph and a source vertex, the
// algorithm computes the distance (named level) of every vertex from the source.
void bfs_simplified_hybrid(graph_t* graph, vid_t source, level_t* levels) {
  // Initialize the global state.
  totem_memset(levels, INF_LEVEL, totem_engine_vertex_count(), TOTEM_MEM_HOST);
  state_g.levels = levels;
  state_g.source = source;

  // Configure and trigger Totem's BSP engine. TOTEM_COMM_PUSH indicates that the
  // communication direction is from the source to the destination vertex of a
  // remote edge, this is in contrast to TOTEM_COMM_PULL which indicates the
  // opposite. The former is used by algorithms in which a vertex pushes a value to
  // update its neighbors (such as BFS), while the latter is used in algorithms
  // where a vertex pulls the state of its neighbors to update its own state (such
  // as PageRank).
  totem_bsp_config_t config = {
    bfs_compute, bfs_scatter, bfs_init, bfs_finalize, bfs_collect, TOTEM_COMM_PUSH
  };
  totem_bsp_config(&config);
  totem_bsp_execute();
}

// The program's main function.
void main() {
  // Load the graph.
  graph_t* graph;
  graph_initialize("/path/to/graph/file", &graph);

  // Initialize Totem. "attr" includes a number of parameters that can be set, the
  // most important of which is the partitioning strategy, which is set to random
  // in TOTEM_DEFAULT_ATTR.
  totem_attr_t attr = TOTEM_DEFAULT_ATTR;
  totem_init(graph, &attr);

  // Allocate the output array and invoke BFS on a random seed.
  level_t* levels = (level_t*)malloc(graph->vertex_count * sizeof(level_t));
  vid_t source = rand() % graph->vertex_count;
  bfs_simplified_hybrid(graph, source, levels);
  graph_finalize(graph);
}
```

**REFERENCES**

AGARWAL, V., PETRINI, F., PASETTO, D., AND BADER, D.A. 2010. Scalable Graph Exploration on Multicore Processors. *The International Conference for High Performance Computing, Networking, Storage, and Analysis*.

BARABÁSI, A.-L. 2003. *Linked: How Everything Is Connected to Everything Else and What It Means*. Plume.

BARABÁSI, A.-L., ALBERT, R., AND JEONG, H. 2000. Scale-Free Characteristics of Random Networks: the Topology of the World-Wide Web. *Physica A: Statistical Mechanics and its Applications 281*, 1-4, 69–77.

BARRETT, R., BERRY, M., CHAN, T.F., ET AL. 1994. *Templates for the Solution of Linear Systems: Building Blocks for Iterative Methods, 2nd Edition*. SIAM.

BARROSO, L.A., DEAN, J., AND HOLZLE, U. 2003. Web Search for a Planet: the Google Cluster Architecture. *IEEE Micro 23*, 2, 22–28.

BEAMER, S., ASANOVIĆ, K., AND PATTERSON, D. 2013. Direction-optimizing breadth-first search. *Scientific Programming 21*, 3, 137–148.

BELLMAN, R. 1958. On a Routing Problem. *Quarterly of Applied Mathematics 16*, 87–90.

BOLDI, P., SANTINI, M., AND VIGNA, S. 2008. A Large Time-Aware Web Graph. *ACM SIGIR Forum 42*, 2, 33–38.

BRANDES, U. 2001. A Faster Algorithm for Betweenness Centrality. *Journal of Mathematical Sociology 25*, 2, 163–177.

CHA, M., HADDADI, H., BENEVENUTO, F., AND GUMMADI, P.K. 2010. Measuring User Influence in Twitter: The Million Follower Fallacy. *International AAAI Conference on Weblogs and Social Media*.

CHAKRABARTI, D., ZHAN, Y., AND FALOUTSOS, C. 2004. R-MAT: A Recursive Model for Graph Mining. *SIAM International Conference on Data Mining*.

CHAMBERLAIN, B.L. 1998. *Graph Partitioning Algorithms for Distributing Workloads of Parallel Computations*. .

CHAMBERS, J.M. 1971. Algorithm 410 Partial Sorting. *Communications of the ACM 14*, 5, 357–358.

CHHUGANI, J., SATISH, N., KIM, C., SEWALL, J., AND DUBEY, P. 2012. Fast and Efficient Graph Traversal Algorithm for CPUs: Maximizing Single-Node Efficiency. *International Parallel and Distributed Processing Symposium*.

CURTISS, M., BECKER, I., BOSMAN, T., ET AL. 2013. Unicorn: A System for Searching the Social Graph. *Proceedings of the VLDB Endowment 6*, 11, 1150–1161.

DAVIDSON, A., BAXTER, S., GARLAND, M., AND OWENS, J.D. 2014. Work-Efficient Parallel GPU Methods for Single-Source Shortest Paths. *Parallel and Distributed Processing Symposium, 2014 IEEE 28th International*, 349–359.

DIJKSTRA, E.W. 1959. A Note on Two Problems in Connexion with Graphs. *NUMERISCHE MATHEMATIK 1*, 1, 269–271.

EDIGER, D., MCCOLL, R., RIEDY, J., AND BADER, D.A. 2012. STINGER: High Performance Data Structure for Streaming Graphs. *High Performance Extreme Computing*.

ERDŐS, P. AND RÉNYI, A. 1960. On the Evolution of Random Graphs. *Publications of tke Matkemafical Insfifufe of the Hungarian Academy of Sciences 5*.

FALOUTSOS, M., FALOUTSOS, P., AND FALOUTSOS, C. 1999. On Power-Law Relationships of the Internet Topology. *ACM SIGCOMM Computer Communication Review 29*, 4, 251–262.

FELDMANN, A. 2012. Fast Balanced Partitioning Is Hard Even on Grids and Trees. In: B. Rovan, V. Sassone and P. Widmayer, eds., *Mathematical Foundations of Computer Science 2012*. Springer Berlin / Heidelberg, 372–382.

FORD, L.A. 1956. Network Flow Theory. *Report P-923*.

GAREY, M.R., JOHNSON, D.S., AND STOCKMEYER, L. 1974. Some Simplified NP-Complete Problems. *Symposium on the Theory of Computing*.

GHARAIBEH, A., BELTRÃO COSTA, L., SANTOS-NETO, E., AND RIPEANU, M. 2012. A Yoke of Oxen and a Thousand Chickens for Heavy Lifting Graph Processing. *International Conference on Parallel Architectures and Compilation Techniques*.

GHARAIBEH, A., COSTA, L.B., SANTOS-NETO, E., AND RIPEANU, M. 2013a. On Graphs, GPUs, and Blind Dating: A Workload to Processor Matchmaking Quest. *International Parallel and Distributed Processing Symposium*.

GHARAIBEH, A., SANTOS-NETO, E., BELTRÃO COSTA, L., AND RIPEANU, M. 2013b. The Energy Case for Graph Processing on Hybrid CPU and GPU Systems. *Workshop on Irregular Applications: Architectures and Algorithm*.

GONZALEZ, J.E., LOW, Y., GU, H., BICKSON, D., AND GUESTRIN, C. 2012. PowerGraph: Distributed Graph-Parallel Computation on Natural Graphs. *Symposium on Operating Systems Design and Implementation*.

GUPTA, P., GOEL, A., LIN, J., SHARMA, A., WANG, D., AND ZADEH, R. 2013. WTF: The Who to Follow Service at Twitter. *International World Wide Web Conference*.

HARISH, P., NARAYANAN, P., ALURU, S., PARASHAR, M., BADRINATH, R., AND PRASANNA, V. 2007. Accelerating Large Graph Algorithms on the GPU Using CUDA. *HiPC*.

HONG, S., KIM, S.K., OGUNTEBI, T., AND OLUKOTUN, K. 2011a. Accelerating CUDA Graph Algorithms at Maximum Warp. *Symposium on Principles and Practice of Parallel Programming*.
HONG, S., OGUNTEBI, T., AND OLUKOTUN, K. 2011b. Efficient Parallel Graph Exploration on Multi-Core CPU and GPU. *International Conference on Parallel Architectures and Compilation Techniques*.
IORI, G., DE MASI, G., PRECUP, O.V., GABBI, G., AND CALDARELLI, G. 2008. A Network Analysis of the Italian Overnight Money Market. *Journal of Economic Dynamics and Control 32*, 1, 259–278.
JEONG, H., MASON, S.P., BARABÁSI, A.L., AND OLTVAI, Z.N. 2001. Lethality and Centrality in Protein Networks. *Nature 411*, 6833, 41–2.
KARYPIS, G. AND KUMAR, V. 1998. A Fast and High Quality Multilevel Scheme for Partitioning Irregular Graphs. *SIAM Journal on Scientific Computing 20*, 1.
KATZ, G.J. AND KIDER JR, J.T. 2008. All-Pairs Shortest-Paths for Large Graphs on the GPU. *Symposium on Graphics Hardware*.
KERNIGHAN, B. 1970. An Efficient Heuristic Procedure for Partitioning Graphs. *The Bell System Technical Journal 49*, 1, 291 – 307.
KWAK, H., LEE, C., PARK, H., AND MOON, S. 2010. What is Twitter, a Social Network or a News Media? *International World Wide Web Conference*.
LEE RODGERS, J. AND NICEWANDER, W.A. 1988. Thirteen Ways to Look at the Correlation Coefficient. *The American Statistician 42*, 1, 59–66.
LI, D. AND BECCHI, M. 2013. Deploying Graph Algorithms on GPUs: An Adaptive Solution. *International Parallel and Distributed Processing Symposium*.
MALEWICZ, G., AUSTERN, M.H., BIK, A.J.C., ET AL. 2010. Pregel: A System for Large-Scale Graph Processing. *SIGMOD International Conference on Management of data*.
MCVOY, L. AND STAELIN, C. 1996. lmbench: Portable Tools for Performance Analysis. *USENIX Annual Technical Conference*.
MERRILL, D., MICHAEL, G., AND GRIMSHAW, A. 2012. Scalable GPU Graph Traversal. *Symposium on Principles and Practice of Parallel Programming*.
MEYER, U. AND SANDERS, P. 2003. Delta-stepping: A Parallelizable Shortest Path Algorithm. *J. Algorithms 49*, 1, 114–152.
NGUYEN, D., LENHARTH, A., AND PINGALI, K. 2013. A Lightweight Infrastructure for Graph Analytics. *Symposium on Operating Systems Principles*.
NVIDIA. 2013. TESLA K20 GPU Active Accelerator Board Specification. .
PAGE, L., BRIN, S., MOTWANI, R., AND WINOGRAD, T. 1999. *The PageRank Citation Ranking: Bringing Order to the Web*. .
PINEDO, M.L. 2012. *Scheduling: Theory, Algorithms, and Systems*. Springer Verlag.
ROWSTRON, A., NARAYANAN, D., DONNELLY, A., O'SHEA, G., AND DOUGLAS, A. 2012. Nobody Ever Got Fired for Using Hadoop on a Cluster. *International Workshop on Hot Topics in Cloud Data Processing*.
SALLINEN, S., BORGES, D., GHARAIBEH, A., AND RIPEANU, M. 2014. Exploring Hybrid Hardware and Data Placement Strategies for the Graph 500 Challenge. *SC*.
SHUN, J. AND BLELLOCH, G.E. 2013. Ligra: A Lightweight Graph Processing Framework for Shared Memory. *Symposium on Principles and Practice of Parallel Programming*.
TIAN, Y., BALMIN, A., CORSTEN, S.A., TATIKONDA, S., AND MCPHERSON, J. 2013. From "think like a vertex" to "think like a graph." *Proceedings of the VLDB Endowment 7*, 3.
VALIANT, L.G. 1990. A Bridging Model for Parallel Computation. *Communications of the ACM 33*, 8, 103–111.
WANG, R., CONRAD, C., AND SHAH, S. 2013. Using Set Cover to Optimize a Large-Scale Low Latency Distributed Graph. *Workshop on Hot Topics in Cloud Computing*.
WANG, X.F. AND CHEN, G. 2003. Complex networks: Small-World, Scale-Free and Beyond. *IEEE Circuits and Systems Magazine 3*, 1, 6–20.